\documentclass[]{aastex631}

\newcommand{\tianshu}[1]{#1}
\newcommand{\adam}[1]{\textcolor{blue}{#1}}

\begin{document}

\title{Neutrino-Driven Winds in Three-Dimensional Core-Collapse Supernova Simulations}

\correspondingauthor{Tianshu Wang}
\email{tianshuw@princeton.edu}
\author[0000-0002-0042-9873]{Tianshu Wang}
\affiliation{Department of Astrophysical Sciences, Princeton University, Princeton, NJ 08544}

\author[0000-0002-3099-5024]{Adam Burrows}
\affiliation{Department of Astrophysical Sciences, Princeton University, Princeton, NJ 08544}

%\collaboration{20}{(AAS Journals Data Editors)}

%% Note that the \and command from previous versions of AASTeX is now
%% depreciated in this version as it is no longer necessary. AASTeX 
%% automatically takes care of all commas and "and"s between authors names.

%% AASTeX 6.31 has the new \collaboration and \nocollaboration commands to
%% provide the collaboration status of a group of authors. These commands 
%% can be used either before or after the list of corresponding authors. The
%% argument for \collaboration is the collaboration identifier. Authors are
%% encouraged to surround collaboration identifiers with ()s. The 
%% \nocollaboration command takes no argument and exists to indicate that
%% the nearby authors are not part of surrounding collaborations.

%% Mark off the abstract in the ``abstract'' environment. 
\begin{abstract}
In this paper, we analyze the neutrino-driven winds that emerge in twelve unprecedentedly long-duration 3D core-collapse supernova simulations done using the code F{\sc{ornax}}. The twelve models cover progenitors with ZAMS mass between 9 and 60 solar masses. In all our models, we see transonic outflows that are at least two times as fast as the surrounding ejecta and that originate generically from a PNS surface atmosphere that is turbulent and rotating. We find that winds are common features of 3D simulations, even if there is anisotropic early infall. We find that the basic dynamical properties of 3D winds behave qualitatively similarly to those inferred in the past using simpler 1D models, but that the shape of the emergent wind can be deformed, very aspherical, and channeled by its environment. The thermal properties of winds for less massive progenitors very approximately recapitulate the 1D stationary solutions, while for more massive progenitors they deviate significantly due to aspherical \adam{accretion}. The $Y_e$ temporal evolution in winds is stochastic, and there can be some neutron-rich phases. Though no strong r-process is seen in any model, a weak r-process can be produced and isotopes up to $^{90}$Zr are synthesized in some models. Finally, we find that there is at most a few percent of a solar mass in the integrated wind component, while the energy carried by the wind itself can be as much as 10$-$20\% of the total explosion energy.
\end{abstract}

%% Keywords should appear after the \end{abstract} command. 
%% The AAS Journals now uses Unified Astronomy Thesaurus concepts:
%% https://astrothesaurus.org
%% You will be asked to selected these concepts during the submission process
%% but this old "keyword" functionality is maintained in case authors want
%% to include these concepts in their preprints.

%% From the front matter, we move on to the body of the paper.
%% Sections are demarcated by \section and \subsection, respectively.
%% Observe the use of the LaTeX \label
%% command after the \subsection to give a symbolic KEY to the
%% subsection for cross-referencing in a \ref command.
%% You can use LaTeX's \ref and \label commands to keep track of
%% cross-references to sections, equations, tables, and figures.
%% That way, if you change the order of any elements, LaTeX will
%% automatically renumber them.
%%
%% We recommend that authors also use the natbib \citep
%% and \citet commands to identify citations.  The citations are
%% tied to the reference list via symbolic KEYs. The KEY corresponds
%% to the KEY in the \bibitem in the reference list below. 

\section{Introduction}

The successful explosion of a core-collapse supernova (CCSN) sweeps away a large fraction of the infalling matter enveloping the proto-neutron star (PNS) \citep{muller2017,stockinger2020,burrows2020,bollig2021}. Therefore, the ram pressure exterior to the PNS decreases progressively with time. This enables the post-explosion emergence of a neutrino-driven wind that expands into the outer layers and eventually catches up with the primary supernova ejecta. Similar to the original model for the Parker solar wind \citep{parker1958} and to analytic predictions by \citet{duncan1986}, the atmosphere of the PNS becomes unstable when its bounding pressure subsides to a level (dependent upon the coeval core neutrino luminosities) sufficient to generate such a secondary outflow powered by neutrino heating via charged-current absorption in the PNS atmosphere \citep{burrows1987, burrows1995}. The wind accelerates and becomes transonic, but when it emerges its detailed properties are functions of model and progenitor specifics. 

The neutrino-driven winds have usually been studied using stationary transonic wind models in spherical symmetry, given a PNS mass and neutrino luminosity \citep{duncan1986,qian1996,otsuki2000,wanajo2001,thompson2001}. These are found approximately to comport with long term one-dimensional core-collapse simulations \citep{hudepohl2010,fischer2012,roberts2012}. Nevertheless, realistic multidimensional CCSN simulations suggest more complex behavior. In 2D, \citet{navo2022} see cone-like winds only towards the southern pole in one of their simulations. In 3D, while \citet{stockinger2020} witness spherical winds for relatively low-mass ZAMS progenitors (e.g., 8.8, 9.0 and 9.6 solar masses), \citet{muller2017} (18 solar masses) and \citet{bollig2021} (17 solar masses), looking for spherically-symmetric outflows, fail to identify them.  We find that this is not due to the absence of the wind, but due to the fact that 1) the wind can emerge seconds after bounce, beyond the simulation time of most researchers, and 2) asymmetrical post-explosion accretion or fallback (later referred to collectively as infall) can interfere with the wind's emergence in some patches of solid-angle around the PNS. This does not mean that the PNS wind does not emerge aspherically, and this is what we see universally using our long-term detailed 3D simulations. This paper presents and details these findings. 

Long-lasting post-explosion accretion is commonly witnessed in many different CCSN models for a variety of progenitors \citep{muller2017,burrows2020,bollig2021}. However, the overall consequences for PNS winds of such long-term infall has to date been unclear. In addition to breaking the spherical symmetry and obstructing some directions, these downflows onto the PNS boost the emergent neutrino luminosity, thus increasing the wind strength along other directions. They also can slightly increase the PNS mass and some of this mass later is ejected in the wind. But the infalling matter can also interact with the expanding wind before the latter reaches a sonic point, thereby thwarting the production of a classic transonic wind. Therefore, it is sometimes hard to determine what the net effect of the asymmetric accretion may be on the total wind intensity without performing time-consuming long-term three-dimensional CCSN simulations. 

The electron fraction ($Y_e$) of the wind material depends on the interaction history with electron-type neutrinos ($\nu_e$) and their anti-particles ($\bar{\nu}_e$). If the wind is neutron-rich ($Y_e<0.5$), it is possible that the rapid neutron-capture process (r-process) can take place. In earlier work \citep{meyer1992,woosley1994}, the wind was thought to have a very high entropy above 300 $k_b$ per nucleon ($k_b$ is the Boltzmann constant) and a strong r-process could take place. However, such high entropies were not produced in subsequent investigations \citep{takahashi1994,qian1996,otsuki2000,thompson2001}. Later work showed that PNS winds were able to produce at most only some of the lightest r-process nuclei \citep{wanajo2013,arcones2013,wanajo2023}, unless other mechanisms are introduced to increase the entropy (e.g., \citet{nevins2023}) or decrease the electron fraction $Y_e$ (e.g., \citet{roberts2012}). All these studies were done in one-dimension, and multi-dimensional effects were ignored. We note that for the same explosion energy, an asymmetrical explosion can manifest faster expansion speeds along the direction of explosion and that matter from slightly deeper PNS layers can be ejected. This can lead to lower $Y_e$ in the ejected matter. Interaction with accreta can also lead to smaller wind-termination radii, and this can extend the duration of nucleosynthesis. 

In this paper, we analyze the early phase of neutrino-driven winds in 12 long-term three-dimensional CCSN simulations. We use multiple methods to prove the existence of winds and to measure their strength. We study the temporal evolution of the physical properties of the wind, and discuss wind nucleosynthesis. We also determine the morphology of the wind regions and the angular mass distribution of the wind ejecta. This paper is arranged as follows: In section \ref{sec:method}, we describe the methods used in this work and summarize the general features of all 12 simulations. In section \ref{sec:existence}, we prove the existence of winds in multiple ways. In Section \ref{sec:evolution}, we study the time-dependent behavior of the winds and in Section \ref{sec:nucleosynthesis} we summarize the nucleosynthesis results. In section \ref{sec:morphology}, we discuss the morphology and direction of the complicated wind structures. Finally, in section \ref{sec:conclusion}, we summarize our results and provide further insights into the 3D PNS wind phenomenon in the core-collapse supernova context.

\section{Method}\label{sec:method}

\begin{figure}
    \centering
    \includegraphics[width=0.49\textwidth]{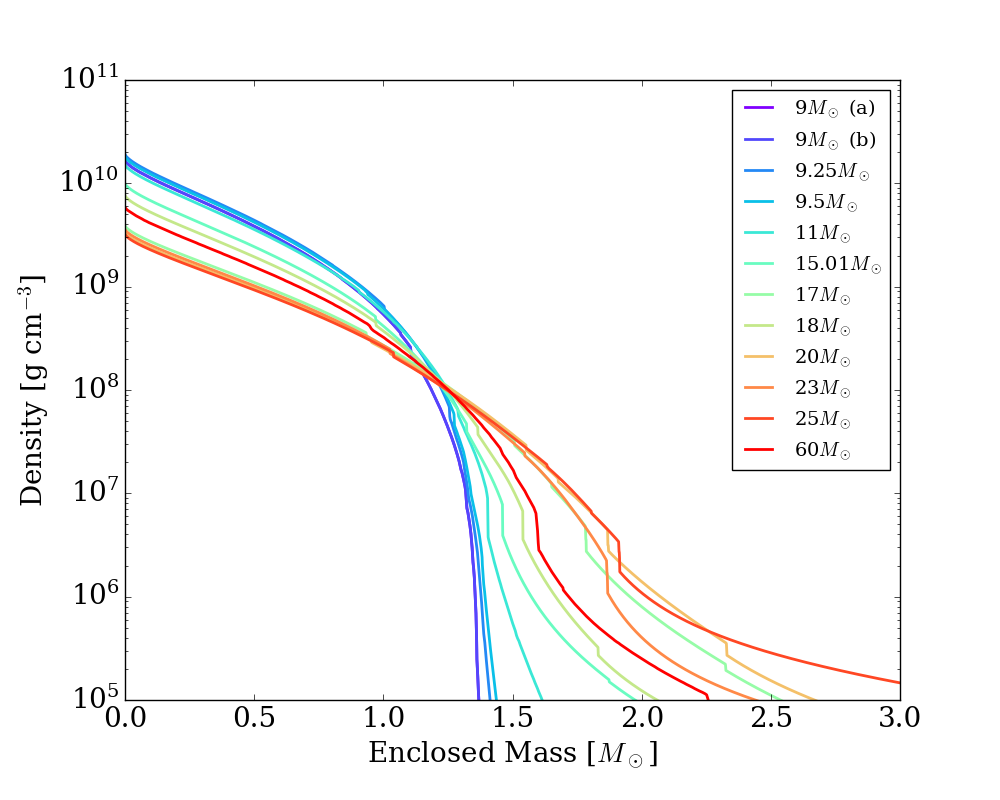}
    \includegraphics[width=0.47\textwidth]{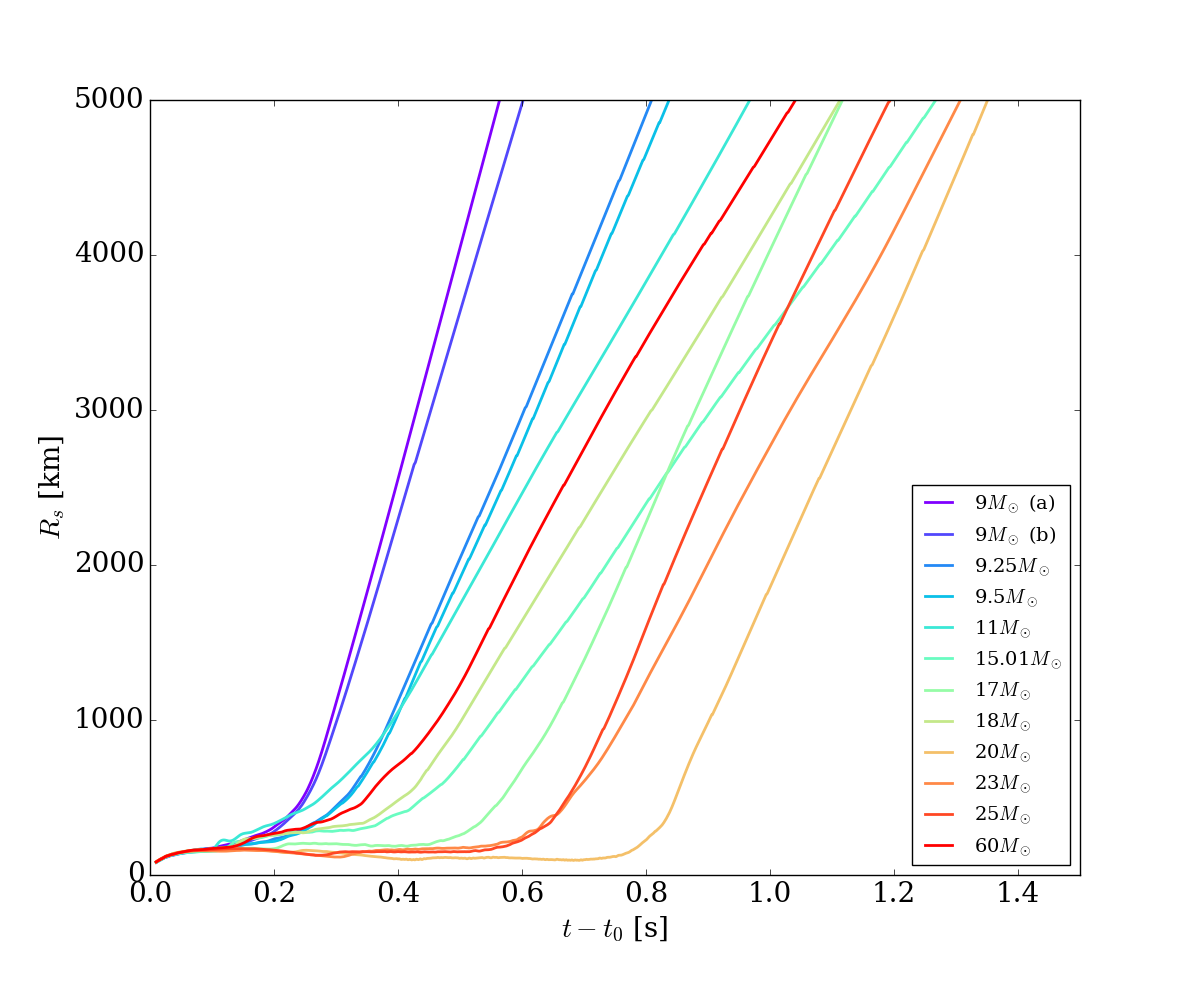}
    \caption{This figure shows the progenitor density profiles (left)  and the shock radii evolution (right). Left panel: the density profiles of all progenitors. The 9, 9.25, 9.5, 11, and 60 $M_\odot$ models are taken from \citet{sukhbold2016} and all other progenitors are from \citet{sukhbold2018}. Right panel: the average shock radii during the first 1.5 seconds. The shock velocities can be found in Table \ref{tab:simulation_summary}.}
    \label{fig:shock}
\end{figure}

For this study, we have used simulations generated by the multi-group multi-dimensional radiation/hydrodynamics code F{\sc{ornax}} \citep{skinner2019,vartanyan2019,burrows2019,burrows2020}. The ZAMS masses of the 12 progenitors are 9 (two models), 9.25, 9.5, 11, 15.01, 17, 18, 20, 23, 25, and 60 $M_{\odot}$. The 9, 9.25, 9.5, 11 and 60 $M_\odot$ progenitors come from \citet{sukhbold2016}, while all the others come from \citet{sukhbold2018}. The initial progenitor density profiles and the evolution of the mean shock radii are shown in Figure \ref{fig:shock}, while some basic properties of the models are summarized in Table \ref{tab:simulation_summary}. None of these models has initial perturbations except for 9(a), which has an initial velocity perturbation between 200-1000 km with $l=10$ and $v_{max}=100$ km/s. We use the SFHo equation of state (EOS) of \citet{steiner2013}, consistent with most known laboratory nuclear physics constraints \citep{tews2017}. All of the models are run with a 1024$\times$128$\times$256 grid, with outer boundary radii varying from 30000 to 100000 km. We use 12 logarithmically-distributed energy groups for each of our three neutrino species (electron-type, anti-electron type, and the rest are bundled as ``$\mu$-type''). 

To follow the movement of wind materials, we add 270,000-300,000 post-processed tracer particles to each simulation. Tracers are passive mass elements that are advected according to the fluid velocity. We use the backward integration method, in which the tracers start at their final positions and the equations of motion are integrated backward in time. \citet{sieverding2022} show that this method leads to more accurate tracer trajectories and thermal histories after the tracers leave the chaotic convection region near the proto-neutron star (PNS). This advantage is essential for studying winds, because wind matter emerges from and through such chaotic regions; using forward integration could lead to wrong ejection times (and, thus, to wrong mass loss rates). The fluid velocity fields of the simulations are saved every millisecond, while the hydrodynamic timesteps of the simulations are around one microsecond. To get better time resolution and to avoid allowing the tracer particles to bypass multiple grid cells in a timestep, we divide the 1-ms time interval into $N_{sub}$ equal length substeps. The velocity field is linearly interpolated in time and space to the position of the tracer particle at each substep. We use adaptive $N_{sub}$ to ensure each tracer no more than half the grid cell size along each direction per substep. This leads to $N_{sub}>100$ when the tracer occupies the chaotic convective regions and $N_{sub}\sim10$ when the tracer is more than a few hundreds of kilometers above the PNS. This adaptive method saves a lot of time taken by the post-processed tracers, because in our long-term simulations the tracers spend most of the time at large radii moving in an almost homologous way. We do not make any assumption in advance concerning the distribution of wind matter, so the tracers are placed logarithmically along the r-direction above 1000 km and uniformly along the $\theta$- and $\phi$-directions at the end of each simulation\footnote{We can do this because we are integrating backward.}.

The nucleosynthesis calculations for the wind matter are done using SkyNet \citep{lippuner2017}, including 1540 isotopes and the JINA Reaclib \citep{cyburt2010} database. We include neutrino interactions with protons and neutrons, but reactions for the $\nu-$process are not included. The detailed neutrino spectra extracted from F{\sc ornax} are then fitted to Fermi-Dirac functions whose parameters are fed into SkyNet, which requires spectra to be in this simplified format. The nuclear statistical equilibrium (NSE) temperature is set at 0.6 MeV ($\sim$7 GK). SkyNet will switch to the NSE evolution mode if the temperature is above this threshold and the strong interaction timescale is shorter than the timescale of density changes \citep{lippuner2017}. We use the $Y_e$ calculated by F{\sc{ornax}} when the temperature is above this threshold, because the neutrino spectra can actually be non-thermal. $Y_e$ evolution below this temperature is handled by SkyNet so that the $\nu p$-process is included.
 
\section{Results}\label{sec:results}

\subsection{Definition of ``Wind"} \label{sec:definition}
Most previous studies on neutrino-driven winds were done in spherical symmetry \citep{duncan1986,qian1996,otsuki2000,wanajo2001,thompson2001,hudepohl2010,fischer2012,roberts2012,wanajo2013,nevins2023} and didn't have multi-dimensional effects. Examples of these 3D effects include convection and turbulent velocity fields, the simultaneous explosion and accretion, and the rotation of the proto-neutron star. These multi-dimensional effects introduce additional complexity and new features into the picture of PNS winds. Therefore, it is essential to clarify the definitions and notations used in this work, since this is the first detailed study of winds using state-of-the-art 3D simulations. 

In this paper, the neutrino-driven wind is defined as a transonic outflow originating from below 80 km, powered by neutrino heating. This definition is based upon the expected dynamical properties of neutrino-driven winds. We have varied the 80-km threshold between 30 and 100 km, and the associated outflows do not much vary. A main consideration is the wind start time, because we wait until the PNS radius falls below this threshold. 
Compared with the winds traditionally studied in 1D, we don't require the wind to be spherical or to be in a (quasi-)steady state. 

The early explosive ejecta lifted by the shock wave (and also driven by neutrino heating) has also experienced a transonic phase, but such ejecta are not considered winds because their interaction with infalling matter leads to very different dynamical and thermal histories. Therefore, it is important to distinguish these two types of outflow. One major difference is the formation time, as the early material is ejected just after the explosion directly by the supernova blast wave, while the winds emerge later on. This time difference is discussed in detail in Section \ref{sec:evolution}.

\begin{figure}
    \centering
    \includegraphics[width=0.48\textwidth]{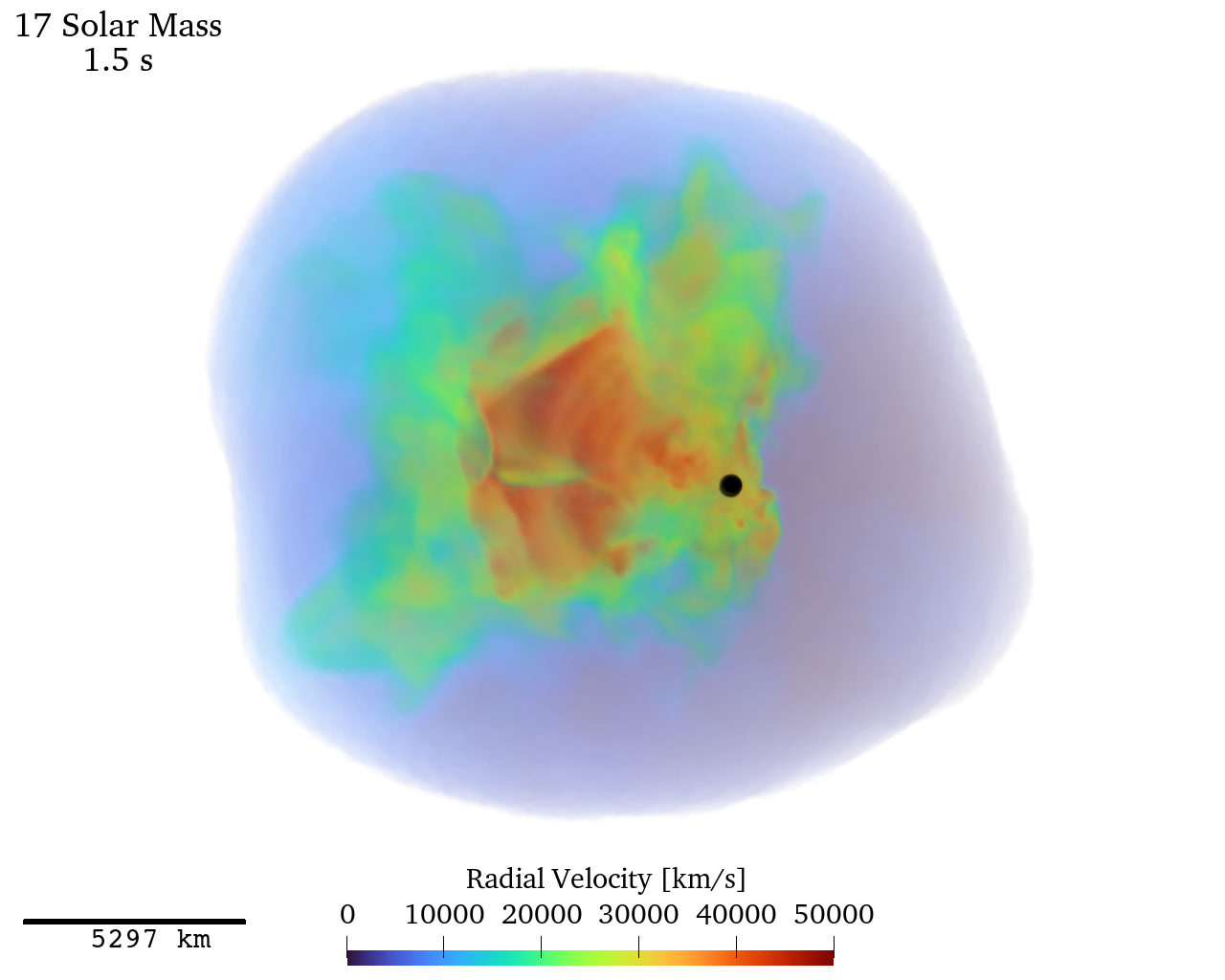}
    \includegraphics[width=0.48\textwidth]{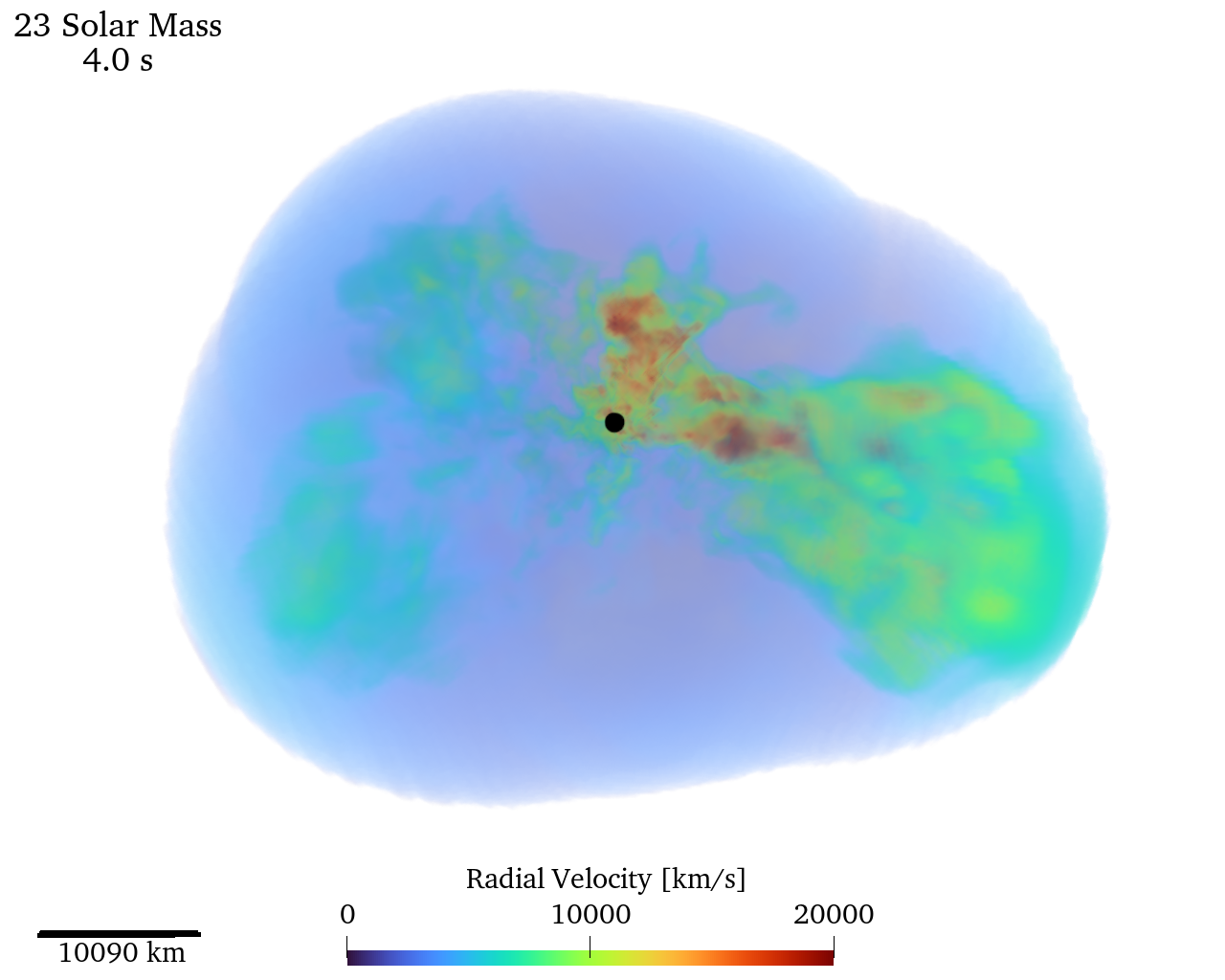}
    \caption{3D volume rendering of the 17 $M_\odot$ model at 1.5 seconds and the 23 $M_\odot$ model at 4.0 seconds after bounce. Colors render the radial velocities (red, higher). The lower velocity regions are made more translucent to show the inner structures. The black dot is the position of the PNS. Left: the 17 $M_\odot$ model. At this time, the shock radius on this plot is about 12000 km and the explosive shock velocity is around 10000 km/s. The secondary shocks, due to the interaction of the wind with the blast ejecta, can be seen clearly on this plot where color changes from dark red to green, meaning that the radial velocity changes from above 40000 km/s to below 20000 km s$^{-1}$. The secondary wind structure is cone-like in this model. Right: the 23 $M_\odot$ model. At this time, the shock radius is about 25000 km from the center and the explosion shock velocity is around 7000 km s$^{-1}$. The opening angle of the secondary structure in this model is significantly smaller, and it is more tube-like and distorted.}
    \label{fig:second-shock}
\end{figure}

Another major difference between winds and early ejecta is the velocity, where winds are usually at least two times faster than the earlier ejecta. This results in wind-termination shocks (also called secondary shocks) behind the explosive shock wave, which can be seen in Figure \ref{fig:second-shock}. In this figure, we show the radial velocity fields of the 17 $M_\odot$ and the 23 $M_\odot$ 3D models. In both models, there are regions with significantly higher velocities than found in surrounding ejecta, and the interface regions are wind-termination shocks.

Such high-velocity regions and wind-termination shocks are general features in our simulations, which means that the wind phenomenon is a common aspect of long-term CCSN simulations. However, there is also some variation. First, the termination velocities in different models vary by a factor of three, ranging from around 15000 km s$^{-1}$ to above 45000 km s$^{-1}$. Even the slowest wind termination velocity is still about two times faster than the surrounding ejecta. Second, the sizes and shapes of the wind regions vary significantly. The 17 $M_\odot$ model is one of our most energetic explosions, and the explosion sweeps away the envelope materials more easily. Thus, its winds are less influenced by accretion and they form a cone-like region along the explosion direction, as indicated in Figure \ref{fig:second-shock}. However, there is more infall in other weaker explosions and the winds can become multiple thin, twisted tubes, like those in the 23 $M_\odot$ model. The shapes we describe here are on larger-scales ($\sim$10000 km), which is the combined result of winds, infall, and the more slowly-moving ejecta. Further discussion of the morphology of the winds can be found in Section \ref{sec:morphology}.

\subsection{Existence} \label{sec:existence}
\begin{figure}
    \centering
    \includegraphics[width=0.48\textwidth]{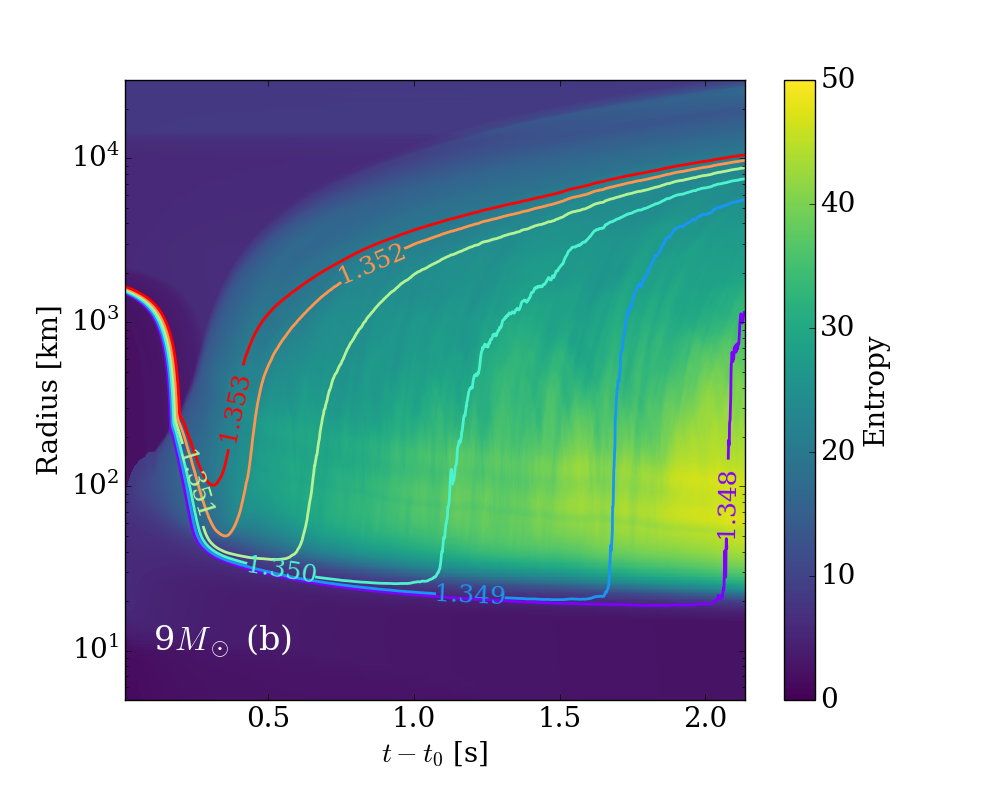}
    \includegraphics[width=0.48\textwidth]{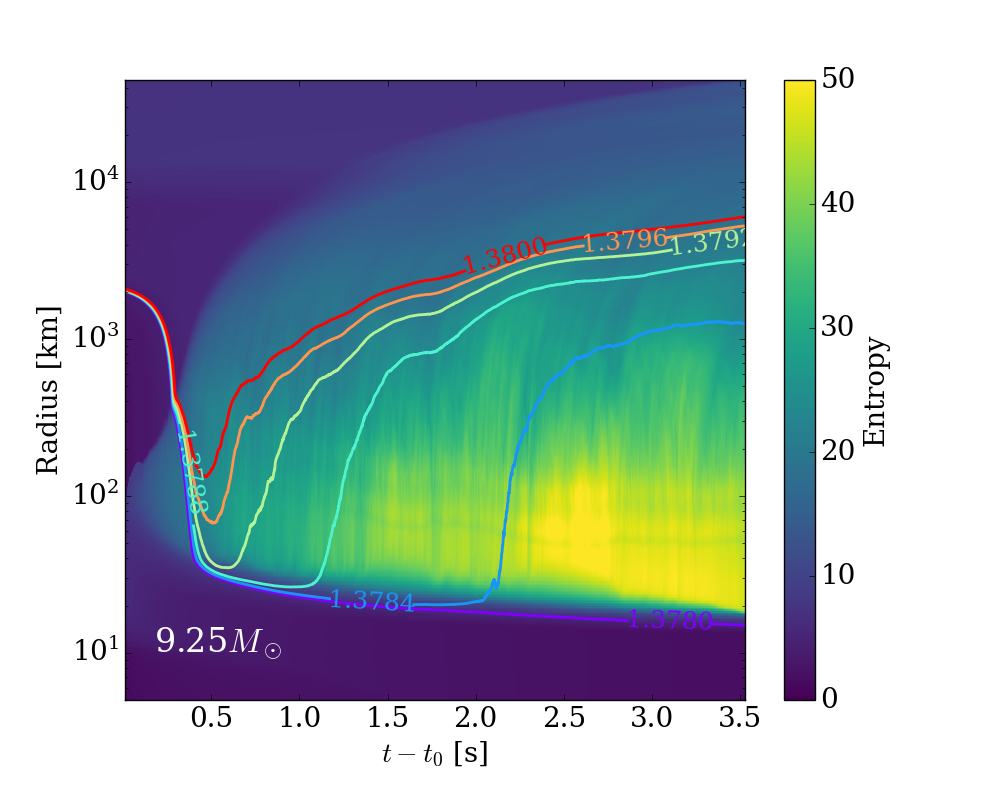}
    \includegraphics[width=0.48\textwidth]{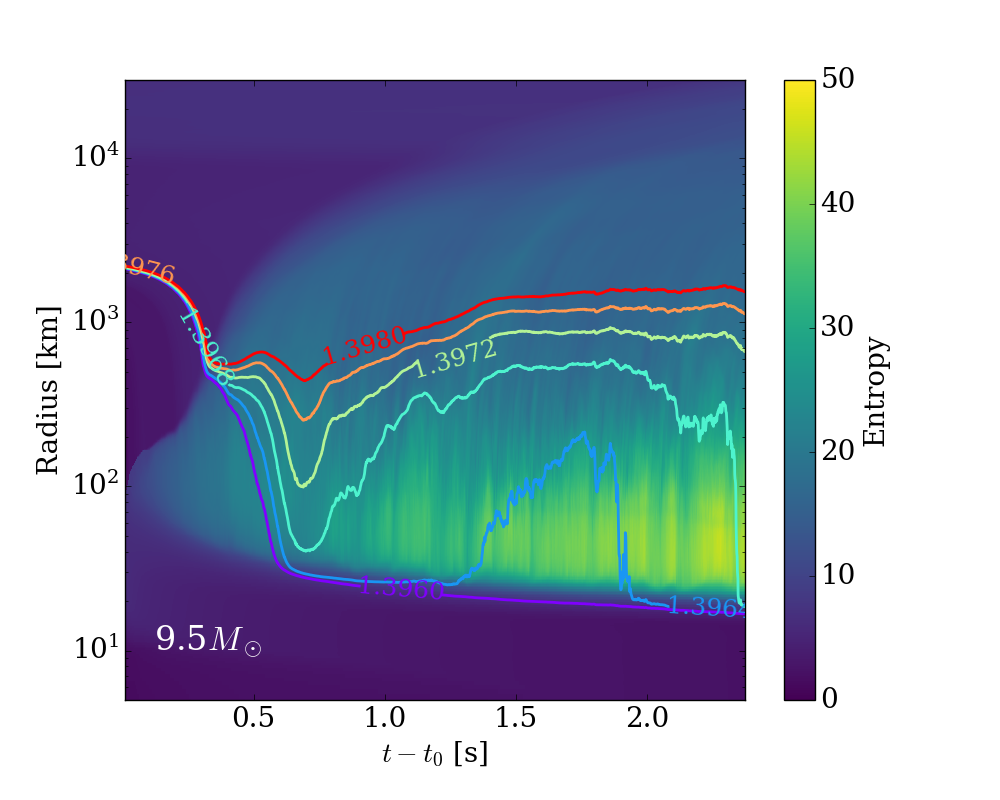}
    \includegraphics[width=0.48\textwidth]{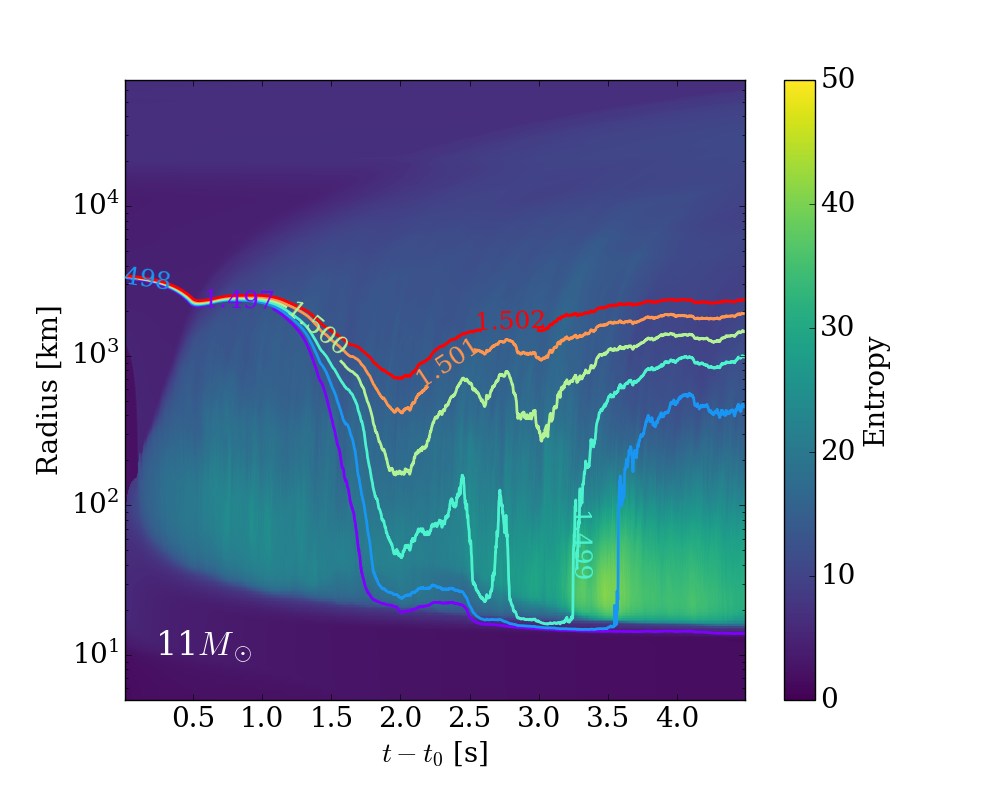}
    \includegraphics[width=0.48\textwidth]{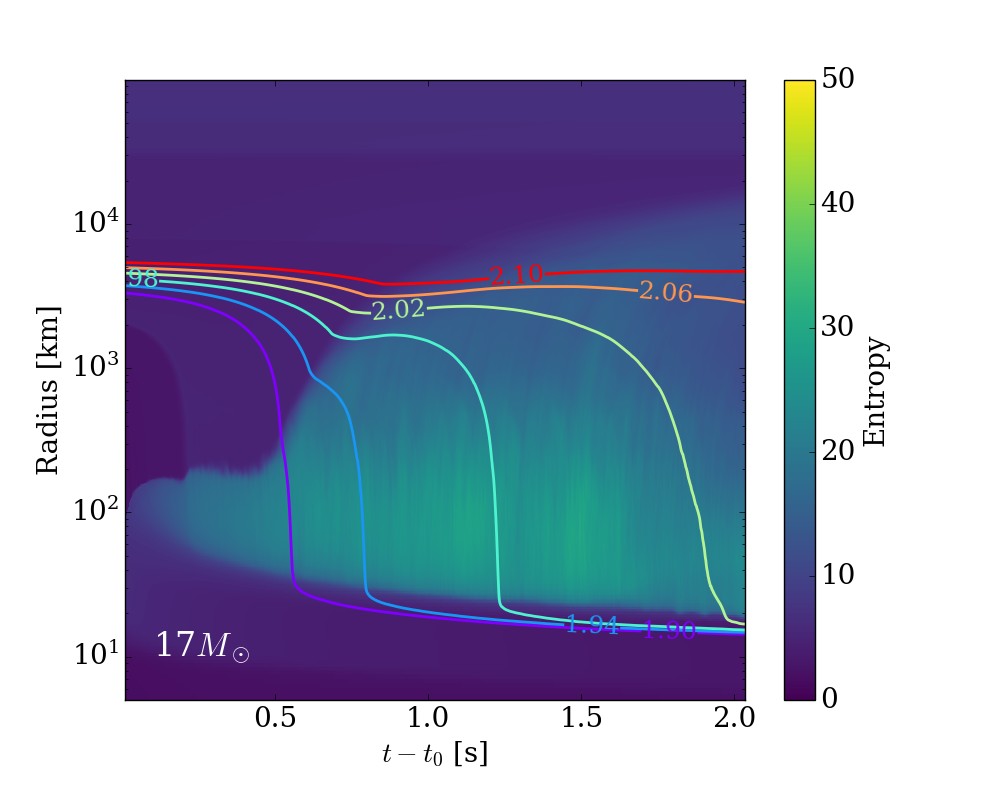}
    \includegraphics[width=0.48\textwidth]{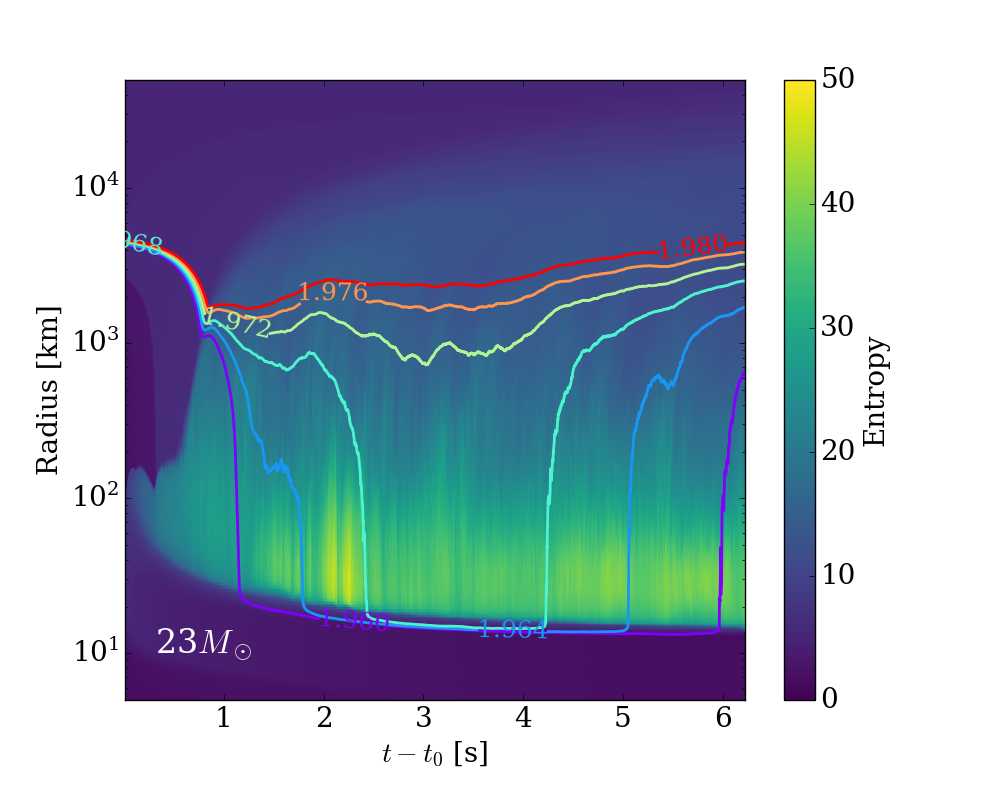}
    \caption{This figure shows the angle-averaged evolution of the 9(b), 9.25, 9.5, 11, 17, and 23 $M_\odot$ models. The lines are radii of constant interior mass and the background color shows the angle-averaged entropy. In the 9(b), 9.25, 11, and 23 $M_\odot$ models, it is clear that the matter falls onto the PNS and stays there for a while before later being ejected; this is a clear behavior of a wind. The 9.5 and 17 $M_\odot$ models don't show such clear behavior, \tianshu{but from Figure \ref{fig:second-shock} we see that there are clear wind regions in the 17 $M_\odot$ model}. \tianshu{This is because the explosions are asymmetrical so the winds and infall can happen simultaneously. The angle-averaged plot can show the wind behavior only when the mass flow rates in winds are stronger than those in accretion, which happens much later, after the emergence of winds. This is the downside of the angle-average method, and it may explain why previous works failed to see spherical winds \citep{muller2017,bollig2021}. Other evidence for winds on these angle-averaged plots are the high-entropy outflows in the background. These high-entropy outflows, starting roughly from the PNS surface, are clearly interpreted as a wind.} } 
    \label{fig:coleman}
\end{figure}

In addition to Figure \ref{fig:second-shock}, in this subsection we provide several different methods to demonstrate the existence of the winds.  
The simplest method is to look at the angle-averaged behavior of the model. Figure \ref{fig:coleman} depicts the radii of constant mass coordinate layers as a function of time. In the 9, 9.25, 11 and 23 $M_\odot$ models, it is clear that the individual layers fall onto the PNS and reside there for a while, but then after a delay move out. This indicates that there are indeed outflows later from the PNS.  Models not manifesting such clear behavior nevertheless experience later-time outflows from the PNS. However, they are weaker and can't be seen so easily in the angle-averaged plots. Nevertheless, the entropy background on these plots indicates that all models have high-entropy outflows from the PNS; this is a clear indication of the universal presence of winds.

A more rigorous way to demonstrate the presence of winds is to follow the matter motion using Lagrangian tracers. We select a ``wind tracer'' based on following criteria: 
\begin{enumerate}
    \item The final radius of the tracer is at least 3000 km.
    \item The tracer has reached a radius below 80 km before its ejection.
    \item The maximum Mach number of the tracer is greater than 1.
    \item The minimum outer shock radius when the specific tracer reaches its smallest radius is at least 3000 km.
\end{enumerate}
The first three conditions ensure that the tracer represents matter in the transonic outflows originating from the PNS. We assume and find that the main acceleration phase of the winds occurs between 100 and 3000 km. This demonstrated to be a good assumption when we check the hydrodynamic histories of the selected tracers. We varied the values used in this assumption, and they lead to similar results. The last condition ensures that the acceleration is not caused by the precursor explosive shock wave itself. As shown in Section \ref{sec:evolution}, this condition is important to technically distinguish the early ejecta and the wind. However, we note that this condition \tianshu{ may miss} its very earliest phase.

\begin{figure}
    \centering
    \includegraphics[width=0.48\textwidth]{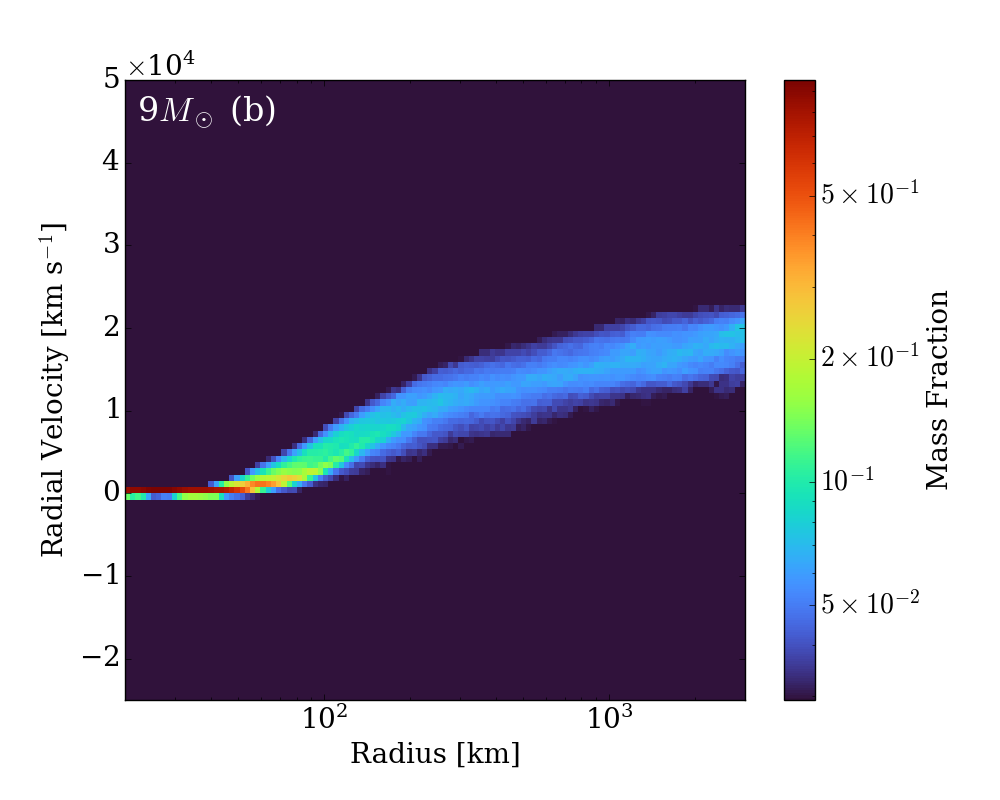}
    \includegraphics[width=0.48\textwidth]{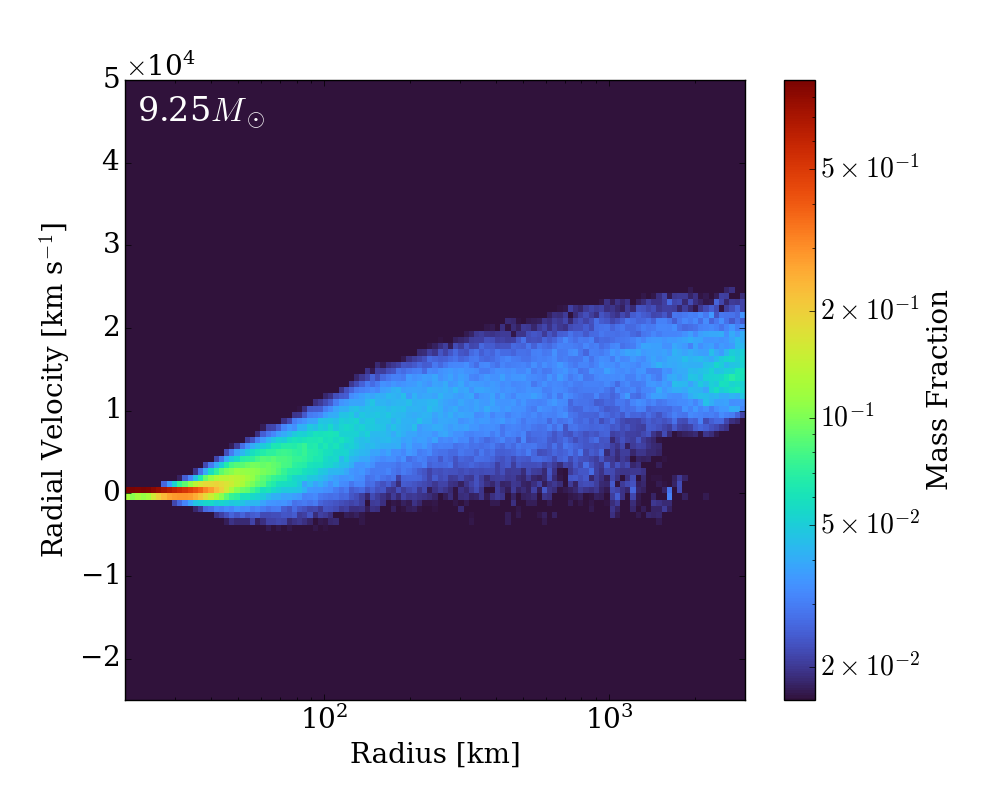}
    \includegraphics[width=0.48\textwidth]{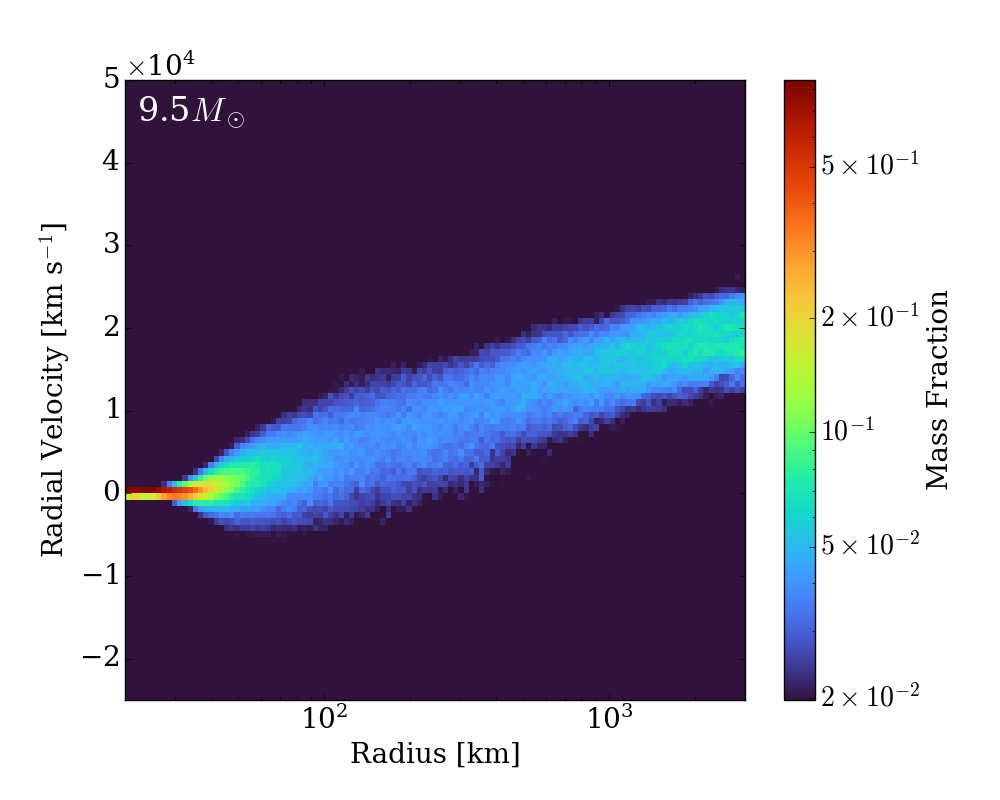}
    \includegraphics[width=0.48\textwidth]{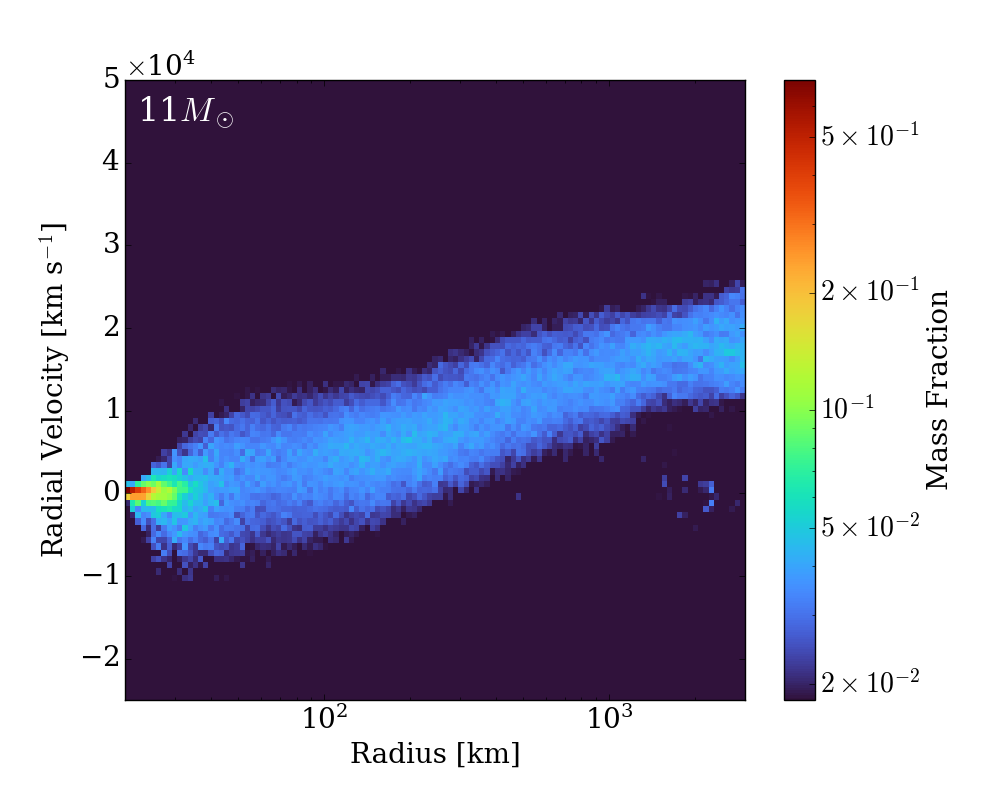}
    \includegraphics[width=0.48\textwidth]{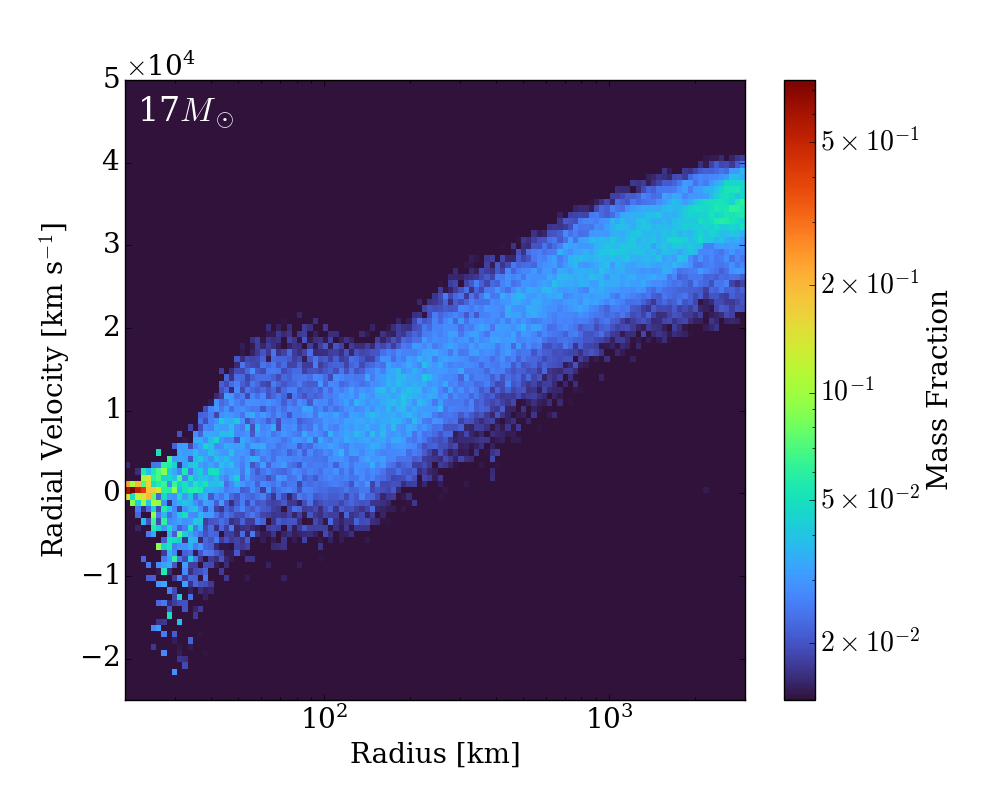}
    \includegraphics[width=0.48\textwidth]{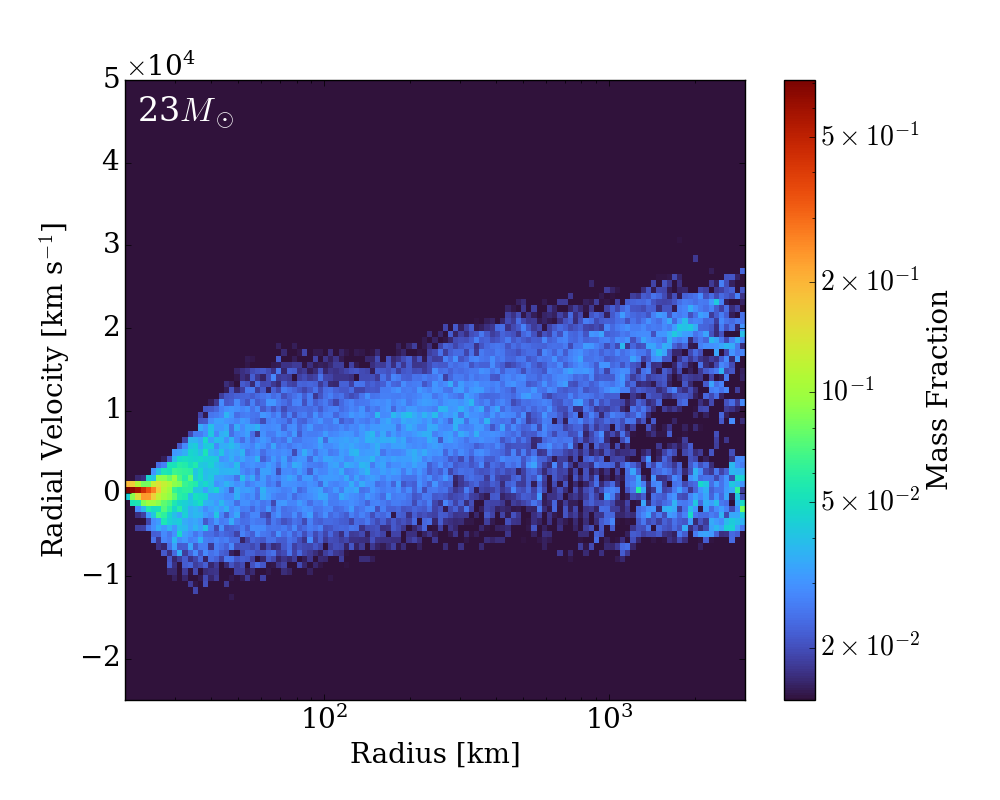}
    \caption{Radial velocity vs. radius histogram of the tracer particles in the acceleration stage. Color shows the mass fraction of all wind material remaining during the acceleration phase at a given radius. Only wind matter is shown here. The tracer trajectories are plotted until they reach their maximum radial velocity, i.e., until they leave the acceleration phase. All models show clear acceleration phases with a similar shape. It can be seen that the termination speed of the wind is 15000-40000 km s$^{-1}$, which is significantly faster than the speed of shock. There is a second band at around zero velocity in some of the models. This is caused by wind matter interacting with infalling matter and pausing its motion for a short period of time, later to be pushed away by the   winds that emerge later. The band shape in the 17$M_\odot$ model looks a bit different from the others because it is a very aspherical explosion; there are two typical wind-termination radii along the explosion and the accretion plume directions.} 
    \label{fig:tracer_vr}
\end{figure}

\begin{figure}
    \centering
    \includegraphics[width=0.48\textwidth]{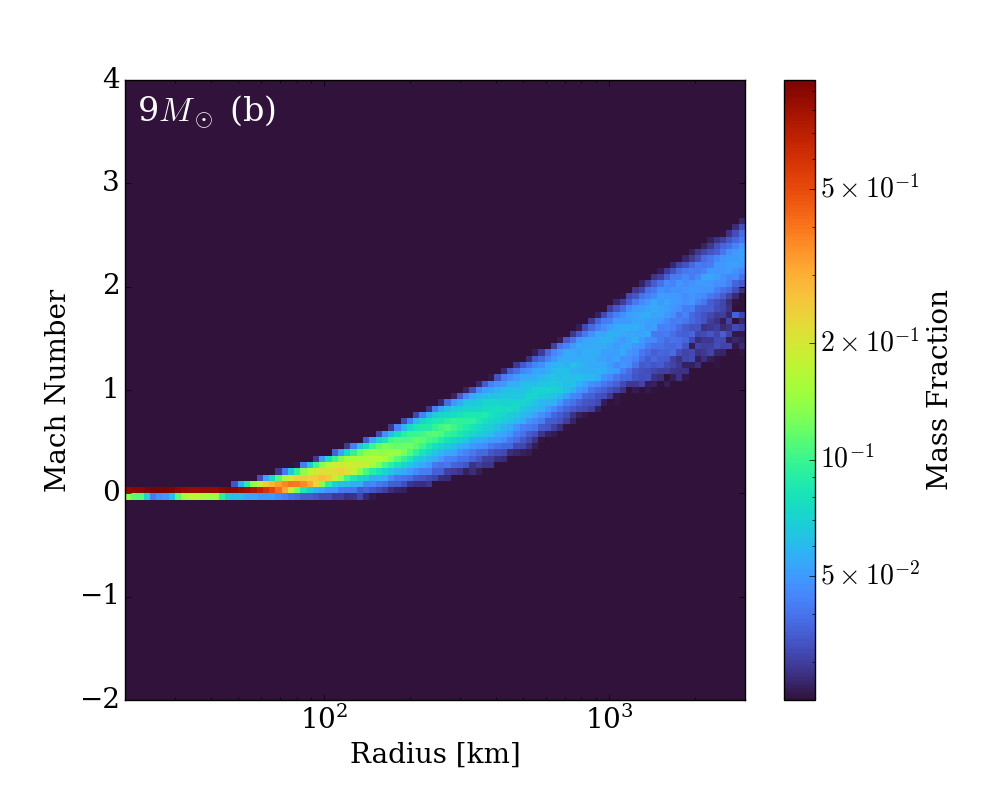}
    \includegraphics[width=0.48\textwidth]{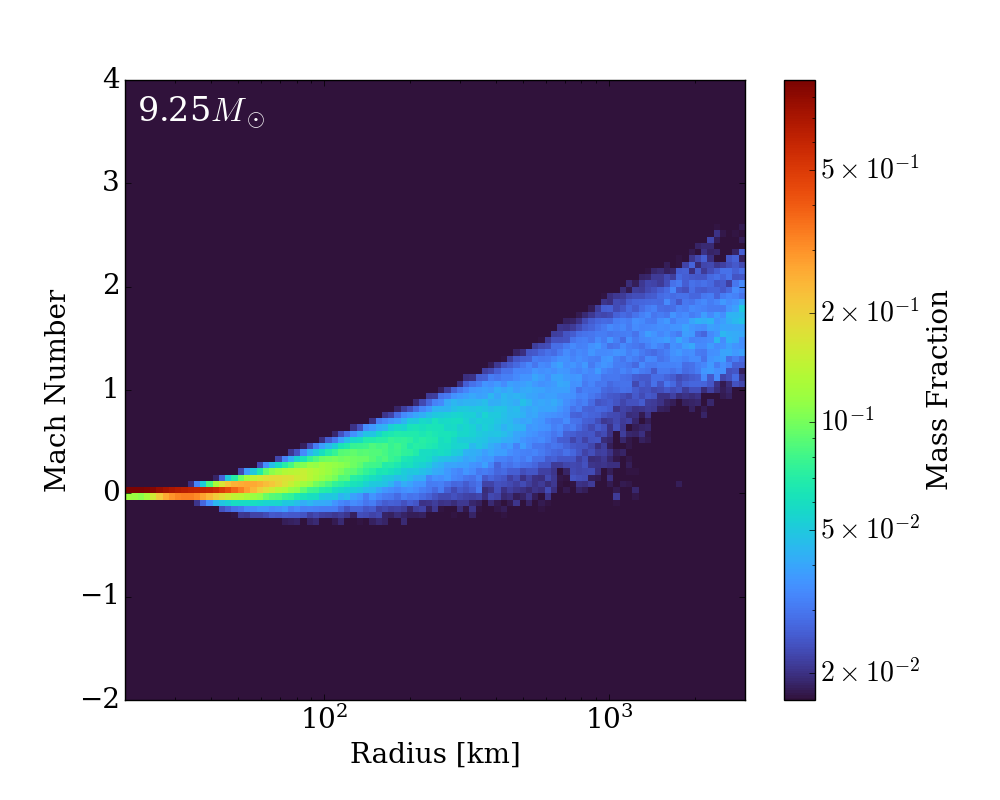}
    \includegraphics[width=0.48\textwidth]{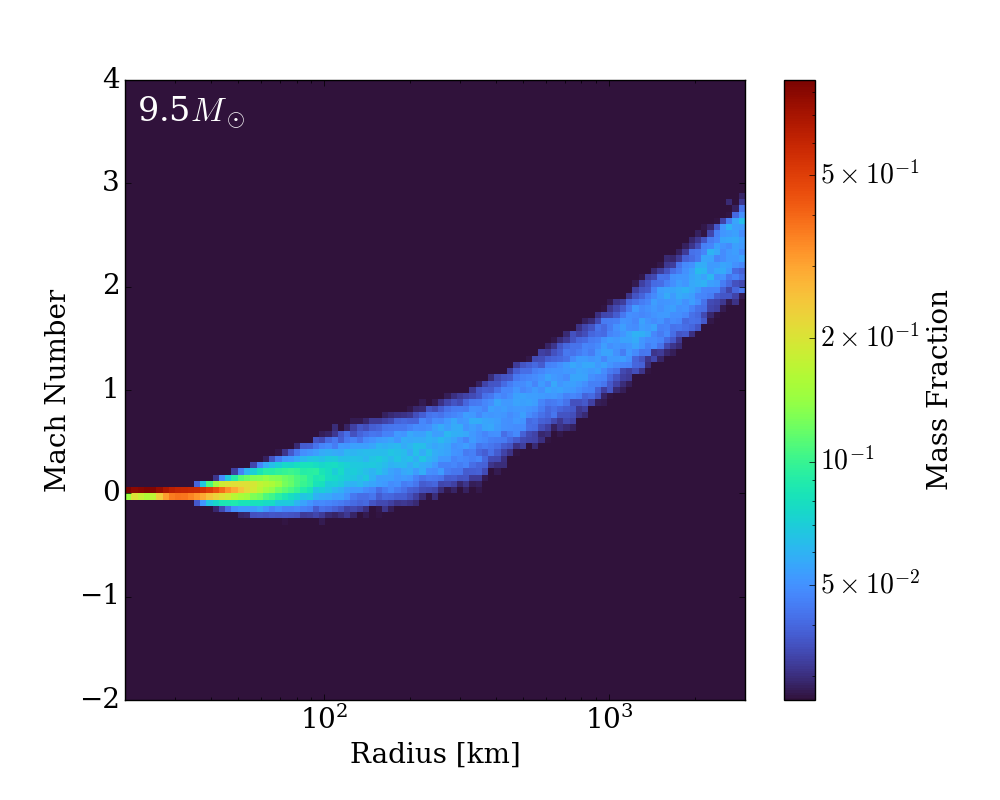}
    \includegraphics[width=0.48\textwidth]{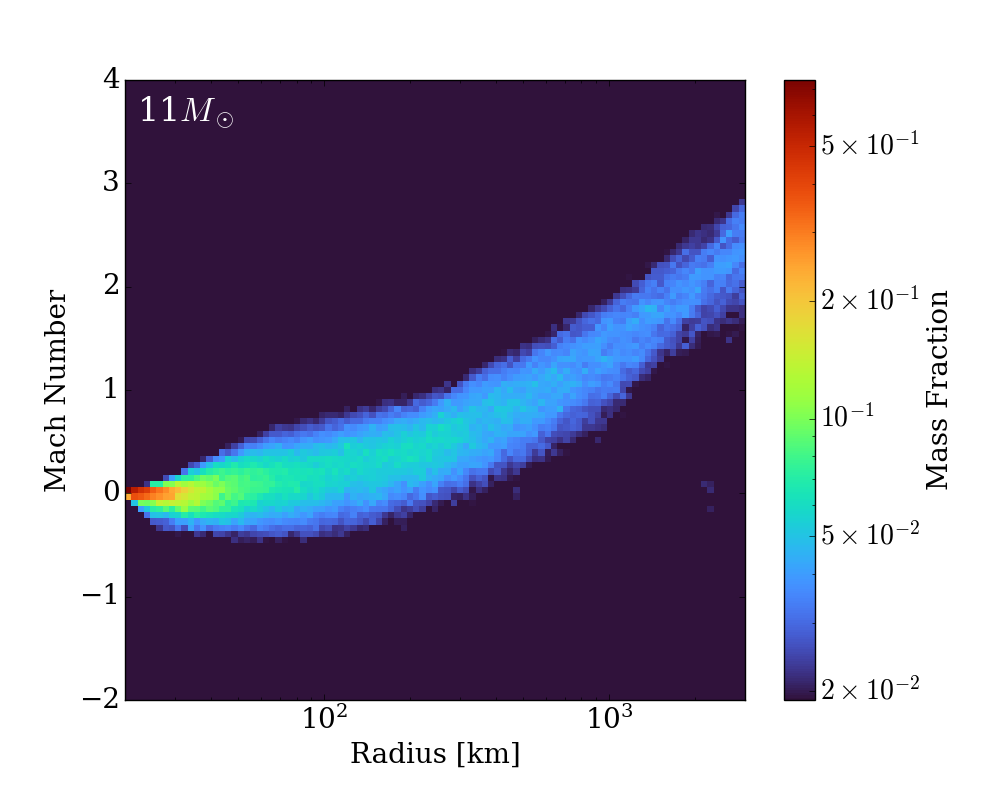}
    \includegraphics[width=0.48\textwidth]{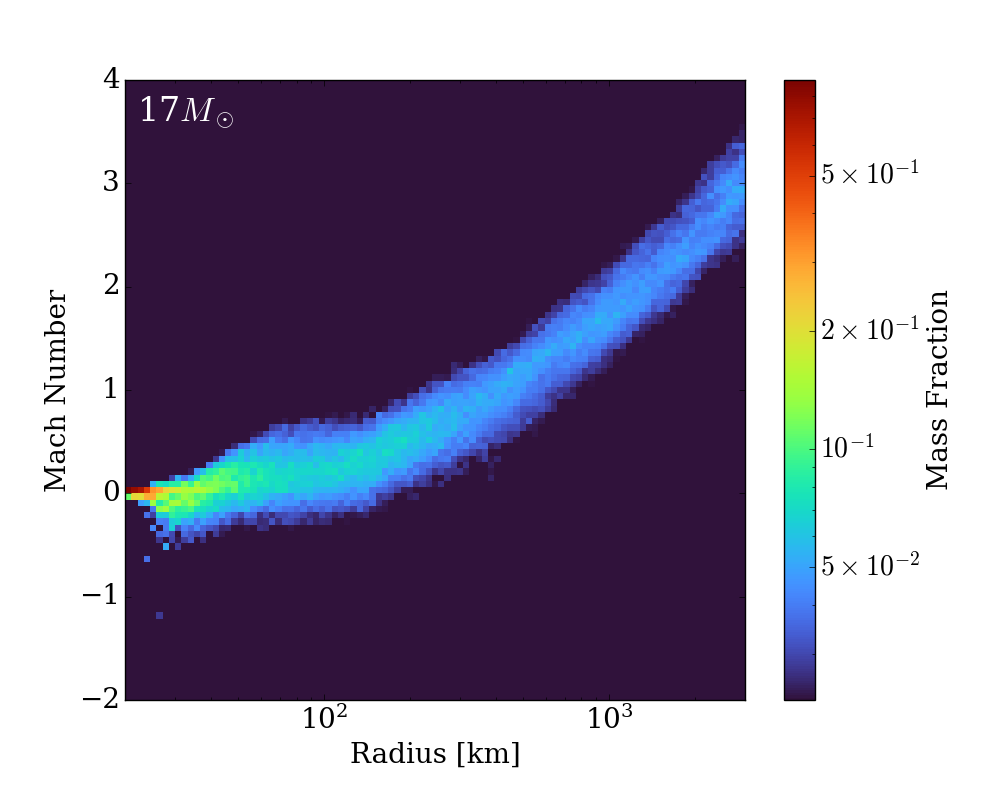}
    \includegraphics[width=0.48\textwidth]{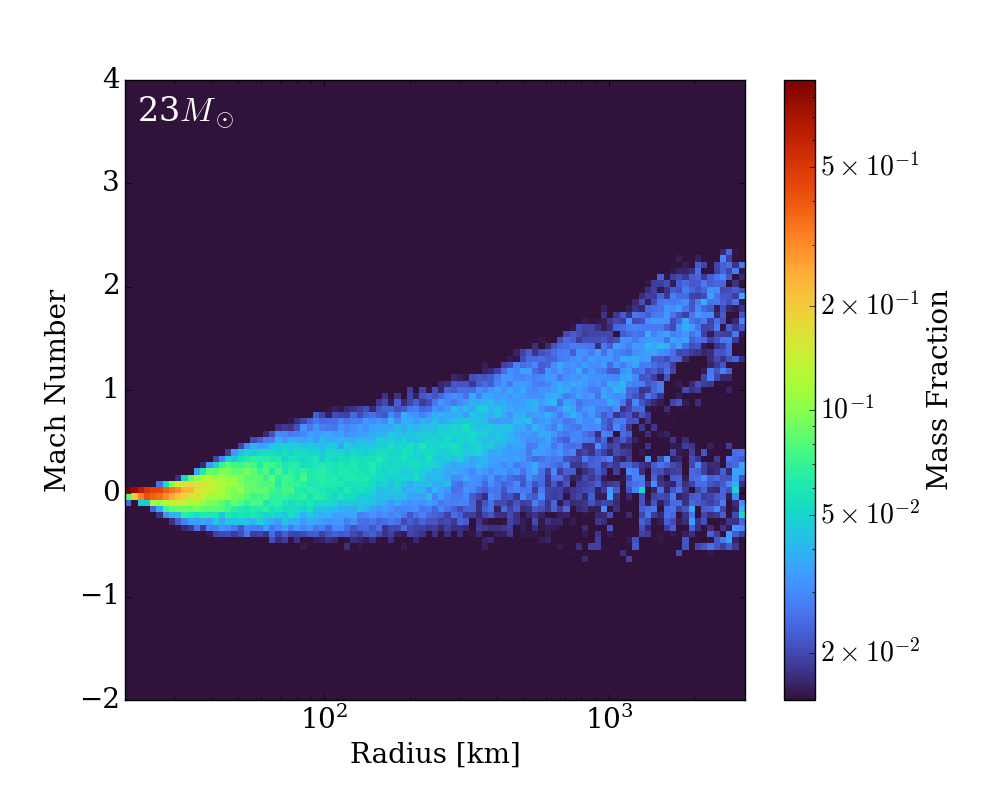}
    \caption{Mach number vs. radius histogram of the tracer particles during the acceleration phase. Color shows the mass fraction of all wind matter still in the acceleration phase at a given radius. Only wind matter is shown here. The tracer trajectories are plotted until they reach their maximum radial velocity. We see transonic bands in all the models, which is a clear signature of a wind. The bands are wider in more massive models due to the stronger accretion plumes they experience. The longer-lasting infall contributes to further turbulence above the PNS and results in outflows having non-zero positive/negative initial radial velocities. The longer-lasting accretion plumes also decelerate the wind outflows and results in wider bands at larger radii. }
    \label{fig:tracer_Ma}
\end{figure}

About 5-10\% (15k-30k) of the tracers we traditionally employ in each simulation satisfy the above wind conditions. This indicates a wind mass of around a few percent of a solar mass, while the energy carried by this component can be 10-20\% of the total supernova explosion energy. Figure \ref{fig:tracer_vr} depicts the relation between velocity and radius for the selected tracers in models 9(b), 9.25, 9.5, 11, 17, and 23 $M_\odot$. The winds are launched from small radii below 80 km and are accelerated to 15000-50000 km s$^{-1}$ at $\sim$3000 km. There is a second branch around zero velocity in some models, and this is because some of the wind matter hits the accreta and stays there for a short period of time. These elements are then pushed away by follow-on winds. Models that explode more weakly seem to have a stronger second branch. Figure \ref{fig:tracer_Ma} shows the wind Mach number as a function of radius and all models clearly show transonic behavior characteristic of winds. 

Apart from the common transonic features, there are certainly differences between models. First, the wind termination velocities vary from 15000 to above 40000 km s$^{-1}$. The termination velocity depends not only on the wind strength, but also the interactions with other debris and continuing infall. Some wind matter exits the acceleration phase earlier because its hits the late-time accreta or more slowly-moving ejecta, and experiences lower termination velocities. This occurs more often in models with stronger accretion and weaker explosions, such as the 23 $M_\odot$ model. Second, the widths of the bands in Figure \ref{fig:tracer_vr} and \ref{fig:tracer_Ma} vary from model to model. This width reflects the velocity variation in wind matter. The turbulent velocity field at smaller radii sets the initial velocity of the wind almost randomly, which sets the width of the band at small radii. \tianshu{Driven by neutrino heating, the turbulent Mach number in this region varies between 0.1 and 1 for different models, and the turbulence extends from the PNS surface to a radius of up to 100 km.} At larger radii, interaction with accreta or slowly-moving blast ejecta also contributes to the depicted variation in the wind velocity. As a result, more massive progenitor models (or models with higher compactness) tend to have greater wind velocity variation because they experience higher initial accretion rates ($\dot{M}$) and stronger turbulence.

\subsection{Evolution} \label{sec:evolution}
\begin{figure}
    \centering
    \includegraphics[width=0.48\textwidth]{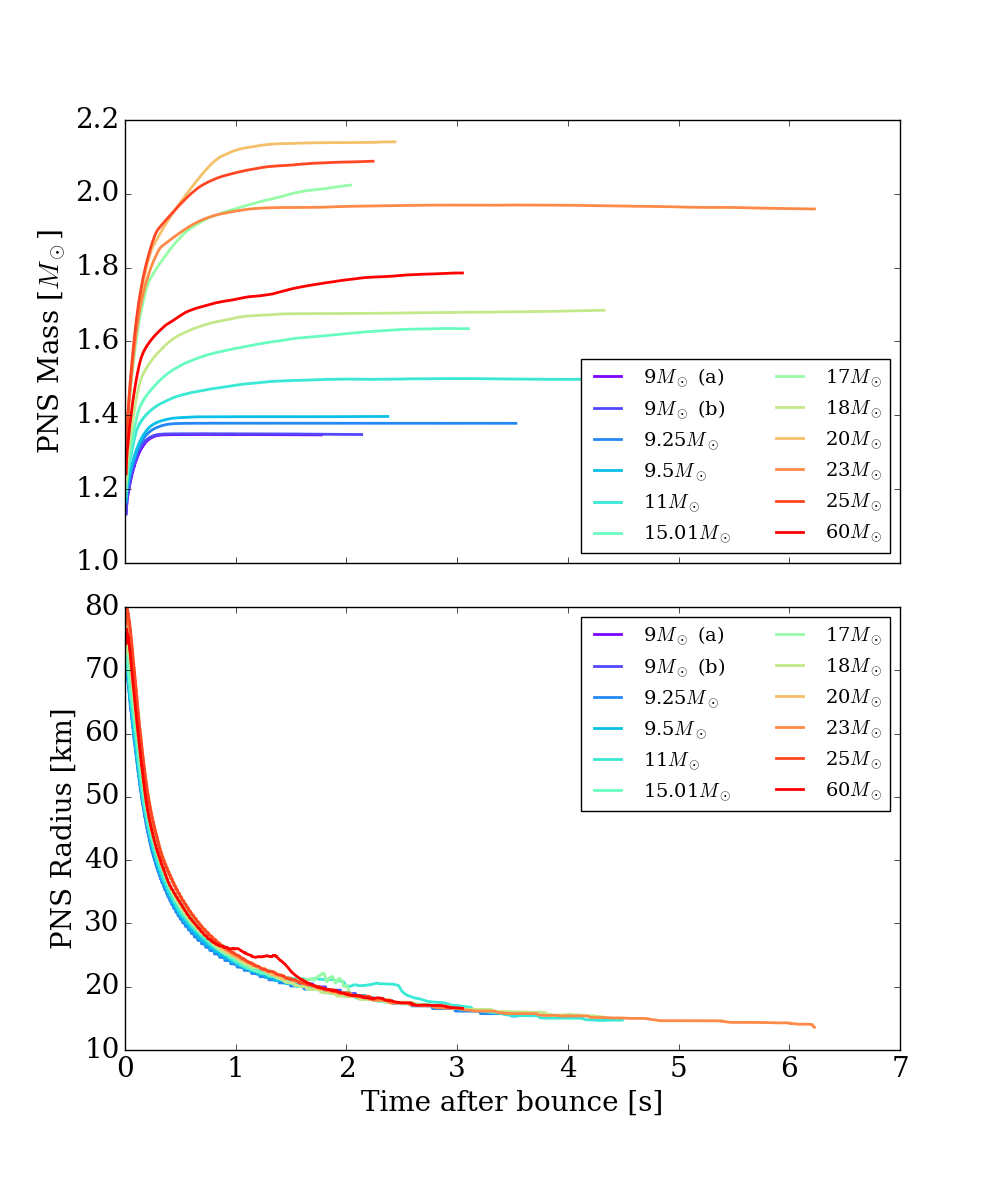}
    \includegraphics[width=0.48\textwidth]{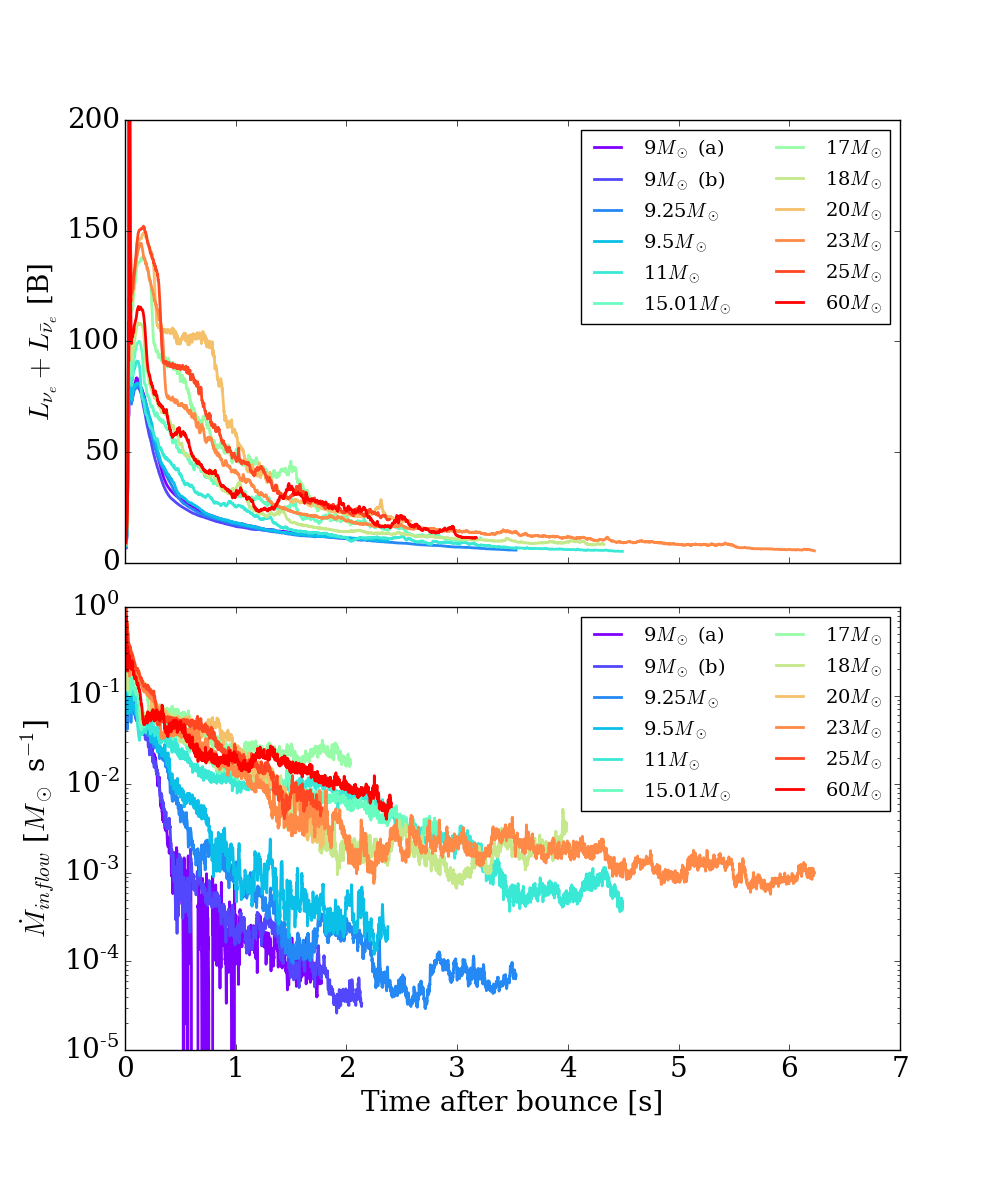}
    \caption{This figure depicts the PNS features, the total luminosities, and the accretion rates for all twelve models. Left panel: average PNS radii and baryonic masses. \tianshu{The PNS surface is defined here to be the $10^{11}$ g cm$^{-3}$ density isosurface.} The PNS radii of all models evolve in a similar way, and the PNS masses stop changing after approximately the first second. Therefore, the PNS properties depicted here don't change much during the wind phase, and they are only secondary factors in the temporal evolution of winds. Right panel: the electron and anti-electron neutrino luminosities measured at 10000 km and the mass inflow rate at 100 km. These are the primary factors that influence the evolution of winds.}
    \label{fig:pns}
\end{figure}

In this subsection we study the temporal evolution of the winds. The left two panels in Figure \ref{fig:pns} shows the PNS mass and radius as a function of time for all the twelve models studied in this paper. The PNS radii of different models decrease at a similar rate, which slows down when the radii are below 20 km, and the accumulated PNS masses stop changing in most models before 1.5 seconds after bounce. Therefore, PNS properties, other than the emergent neutrino luminosities, are only secondary factors in the temporal evolution of winds. The right two panels in Figure \ref{fig:pns} show the neutrino luminosity ($\nu_e+\bar{\nu}_e$) measured at 10000 km and the mass flow rate of the inflows measured at 100 km. Note that this is not the net accretion rate (which is the mass flow rate difference between inflow and outflow). The inflow rates in less massive progenitor models decrease faster, while infall in more massive models generally lasts longer. In the initially non-rotating context, the luminosity and mass accretion rate are the main factors that influence the evolution of winds.

\begin{figure}
    \centering
    \includegraphics[width=0.45\textwidth]{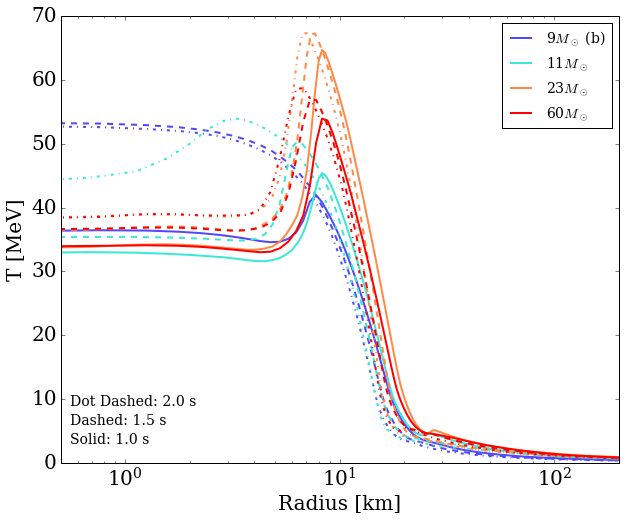}
    \includegraphics[width=0.51\textwidth]{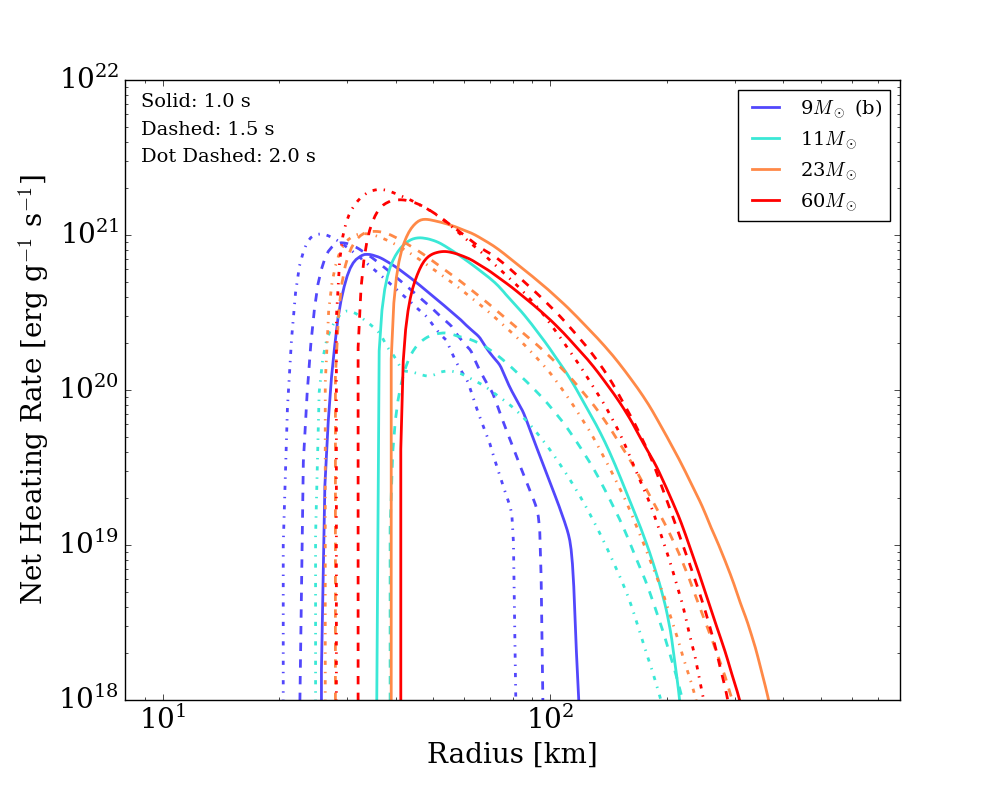}
    \caption{Angle-averaged temperature and heating rate profiles at different times for the 9 $M_\odot$ (b), 11 M$_{\odot}$,  23 $M_\odot$, and 60 M$_{\odot}$ models. Left: Temperature profiles. At early times, the temperature profile has a peak at around 10 km. This peak moves inward as the PNS shrinks, and at the same time the central PNS temperature increases. Right: Neutrino heating rate profiles. As time evolves, the gain region moves to smaller radii. At around $\dot{q}=2.5\times10^{20}$ erg s$^{-1}$, the slope changes due to recombination of free nucleons into $\alpha$ particles. The change of slope can be seen more clearly in the 9(b) model because it is more spherical, while in the other models the smooth profiles seen are in part due to the angle-averaging.}
    \label{fig:qdot}
\end{figure}

The left panel of Figure \ref{fig:qdot} shows the angle-averaged temperatures of the 9(b), 11, 17 and the 23 $M_\odot$ models at 1.0, 1.5, and 2.0 seconds after bounce. At early times, the temperature profile of the PNS has a peak around 10 km. This peak moves inward and diminishes with time. Eventually, the thermal profile becomes similar to the functional form often used to describe the PNS \citep{kaplan2014}. However, achieving this state takes more than a few seconds, and the early phase of the neutrino-driven wind is certainly influenced by the presence of this thermal spike. The right panel of Figure \ref{fig:qdot} depicts the angle-averaged net neutrino heating rate profiles of the same models. The gain region (where matter gains energy from neutrinos) can be clearly seen. In general, the inner boundary of the gain region is about two times the PNS radius, and moves inward as the PNS shrinks. This means that the origination points of the winds also move inward. However, the effect of the velocity spread of the wind matter shown in Figure \ref{fig:tracer_vr} and \ref{fig:tracer_Ma} is more pronounced than this start-point variation. Moreover, on the velocity-radius and Ma-radius plots we don't see a clear temporal change in the band widths and magnitude. This means we can assume that the winds experience roughly the same acceleration phase in the first few seconds. Therefore, the dynamical history of the wind matter is determined only by the wind-termination time (or equivalently, the wind termination radius). 

\begin{figure}
    \centering
    \includegraphics[width=0.48\textwidth]{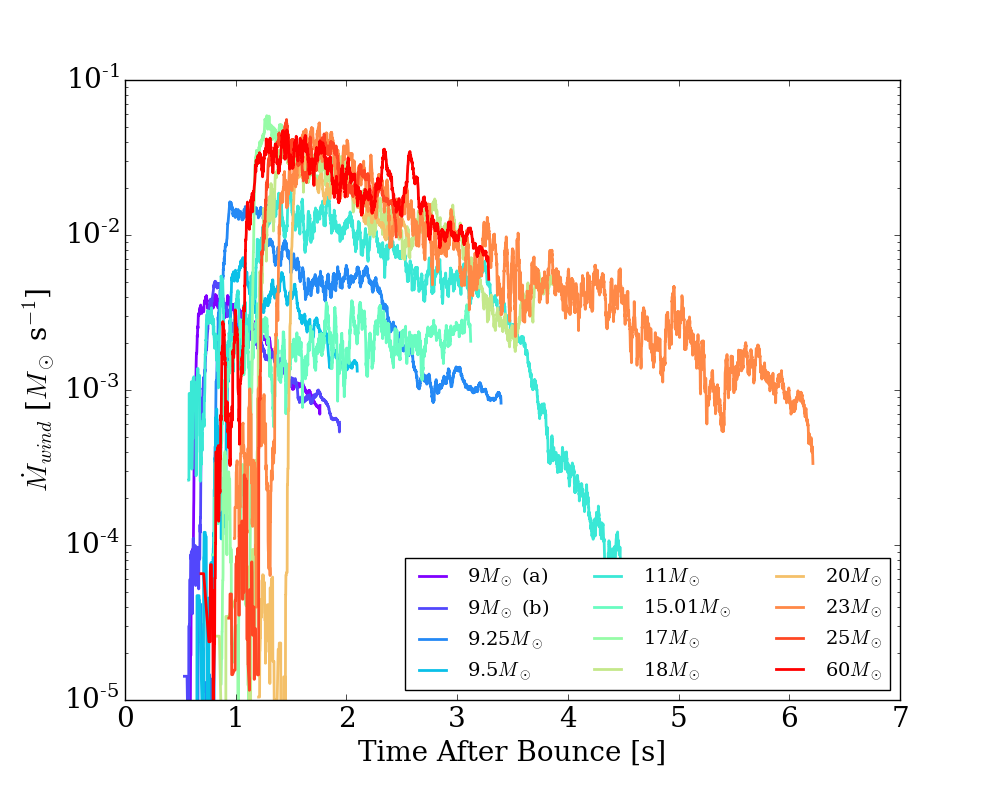}
    \includegraphics[width=0.48\textwidth]{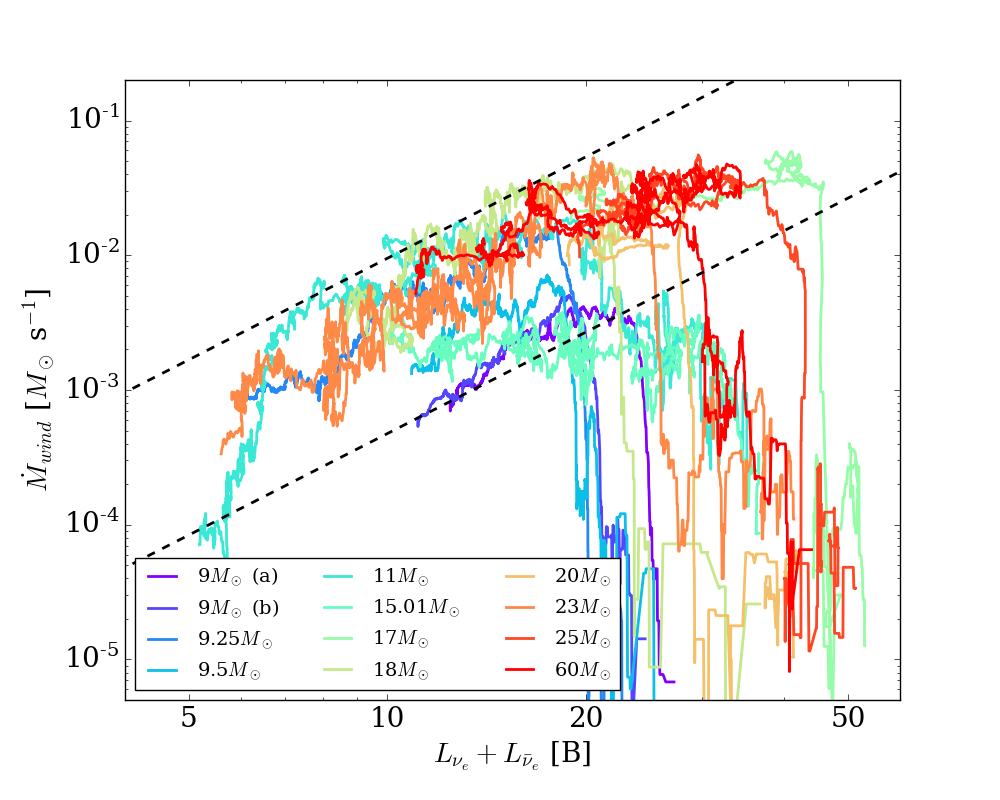}
    \caption{The wind mass flow rates at 3000 km calculated by the tracer method as a function of time (left) and versus luminosity (right) at 10000 km for all twelve models of this study. The mass flow rates and luminosities are both averaged over 25 ms. Left: the wind mass outflow rates as a function of time. It is seen that the outflow rates for different models follow a similar decay pattern, even though their accretion rates (see Figure \ref{fig:pns}) are very different. Right: wind mass outflow rate vs. luminosity. All models evolve from right to left (high luminosity to low luminosity). The black dashed lines provide the slope of $\dot{M}_{wind}\propto L^{2.5}$ relation. It is clear that all models approximately follow the $L^{2.5}$ relation predicted by the spherical stationary wind solutions (e.g., \citet{duncan1986,burrows1987} (with the extra assumption that $L\propto T_{\nu}^4$). However, the other predicted relation $\dot{M}_{wind}\propto M_{pns}^{-2}$ is not seen here, since models with higher PNS masses (such as the 23 $M_\odot$ model) have stronger wind outflow rates than those with lower PNS masses (such as both 9 $M_\odot$ models)}
    \label{fig:mdot-L}
\end{figure}

Figure \ref{fig:mdot-L} shows the temporal evolution of the wind mass flow rates and the relation between the wind mass flow rate and the neutrino luminosity for all the twelve models. The wind mass flow rate is measured at 3000 km using the wind tracers. Despite the very different accretion rates (see Figure \ref{fig:pns}), wind mass flux in all models seems to decay at a roughly similar rate. The peak mass flux of the winds is between $5\times10^{-3}$ and $5\times10^{-2}$$M_\odot$s$^{-1}$. In addition, this mass flux in all models follows the $\dot{M}_{wind}\propto L^{2.5}$ relation (black dashed lines). This $\dot{M}_{wind}\propto L^{2.5}$ power law relation is predicted by the spherical stationary wind solutions \citep{duncan1986,burrows1987,qian1996,thompson2001} (assuming ``$L\propto T^4$," see these papers). However, the predicted dependence on the PNS mass ($\dot{M}_{wind}\propto M_{PNS}^{-2}$) is not seen here, since models with higher PNS mass (like the 17 $M_\odot$ model) don't show weaker winds. It is possible that models with higher PNS masses also have longer-lasting accretion which provides more matter into the wind region and thereby enhances the wind mass loss flux. It is worth mentioning that the wind mass flow rate in some models is only a small fraction of the total mass outflow rate. 

\begin{figure}
    \centering
    \includegraphics[width=0.48\textwidth]{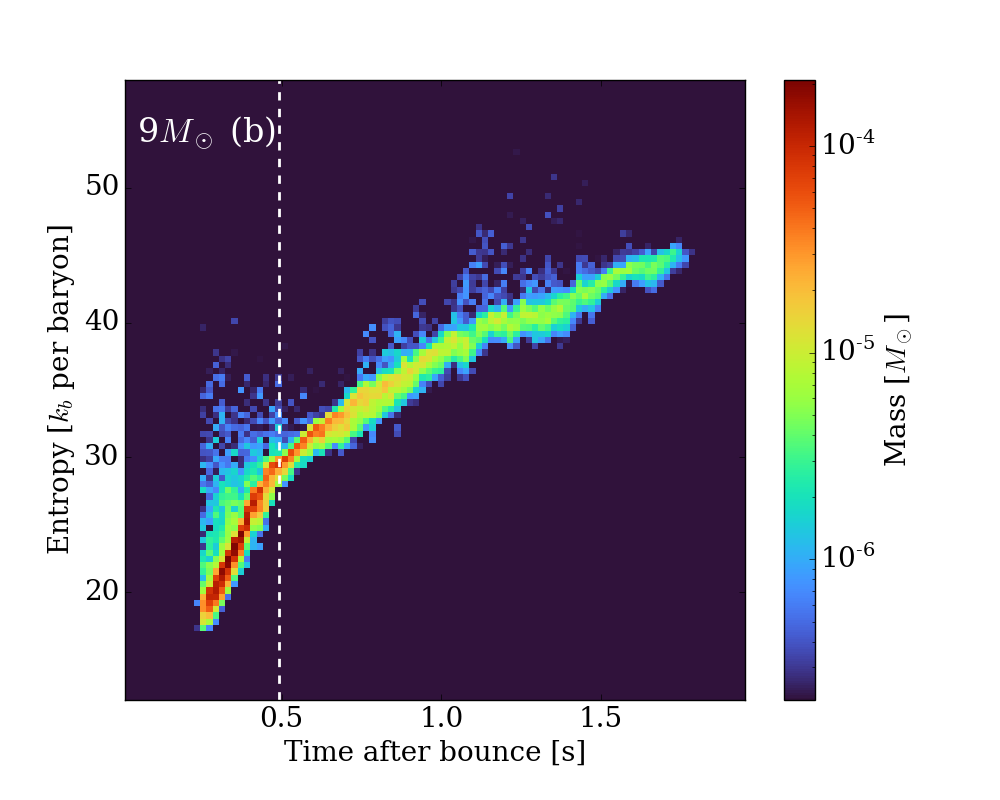}
    \includegraphics[width=0.48\textwidth]{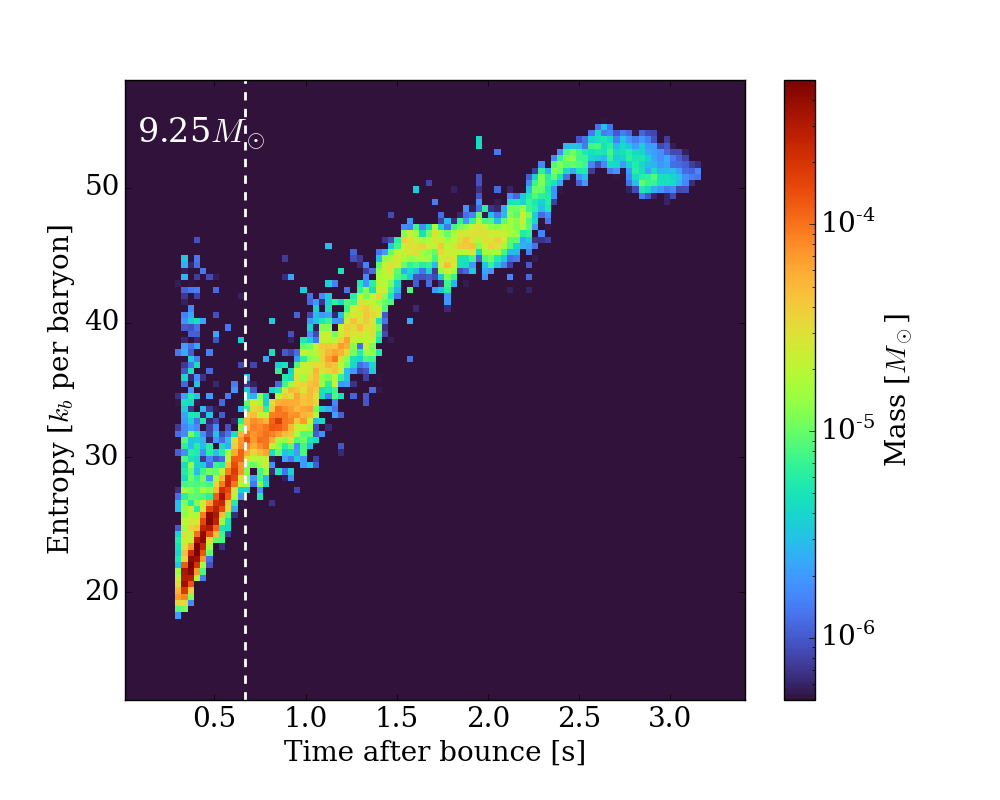}
    \includegraphics[width=0.48\textwidth]{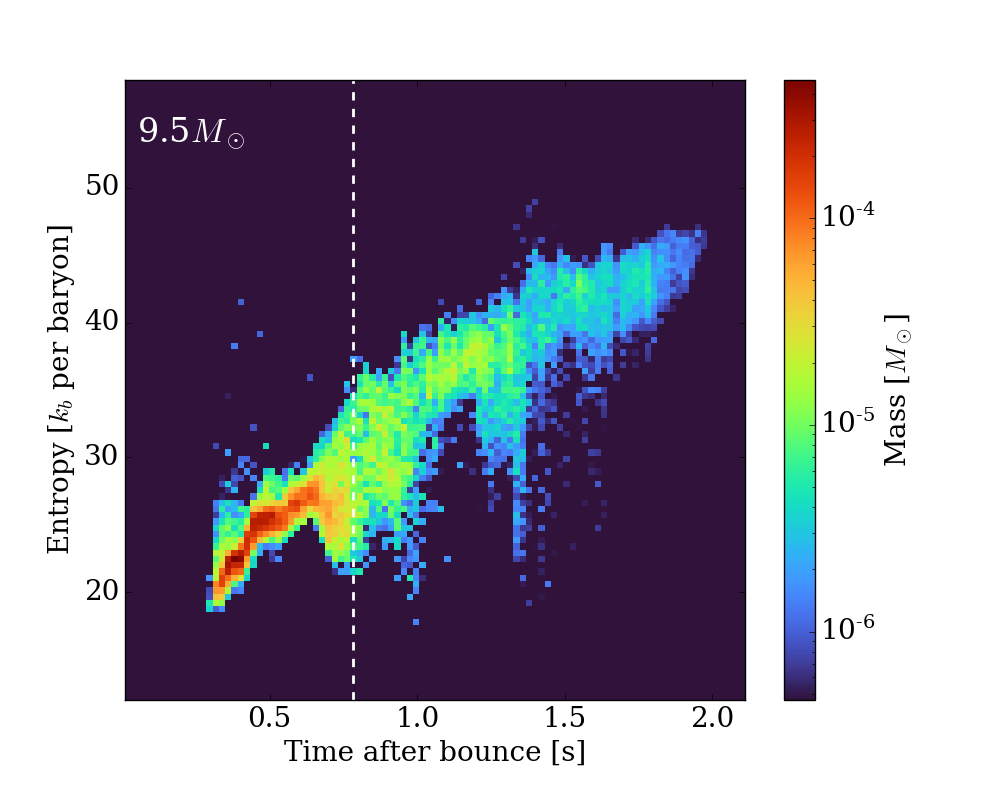}
    \includegraphics[width=0.48\textwidth]{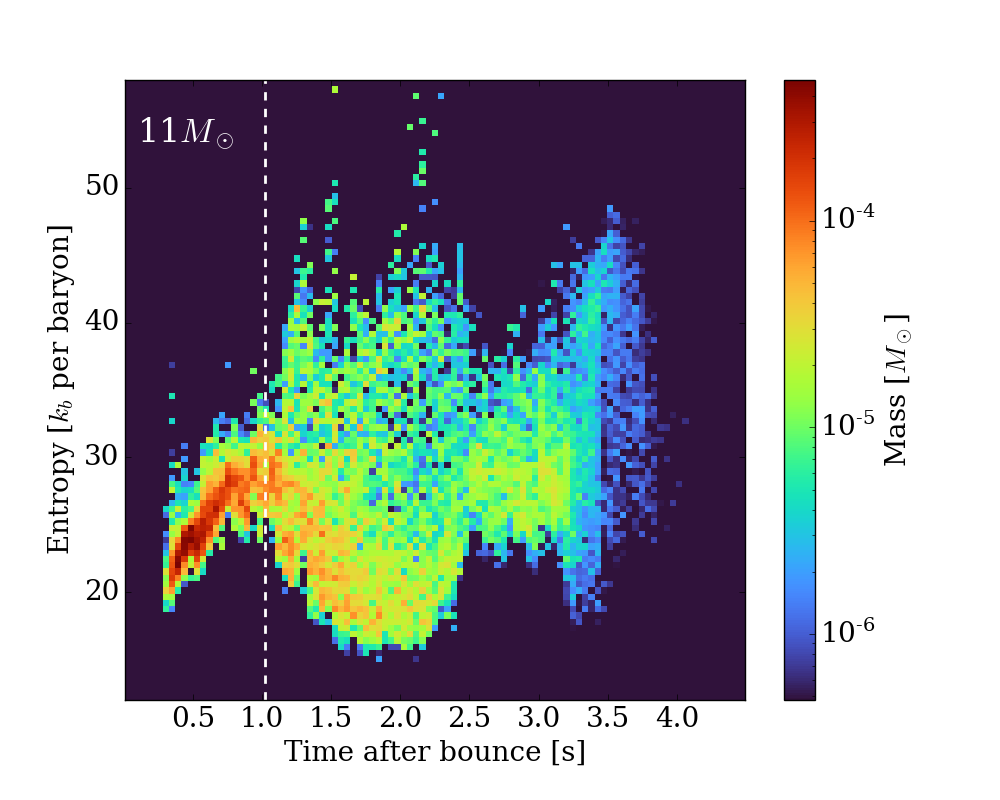}
    \includegraphics[width=0.48\textwidth]{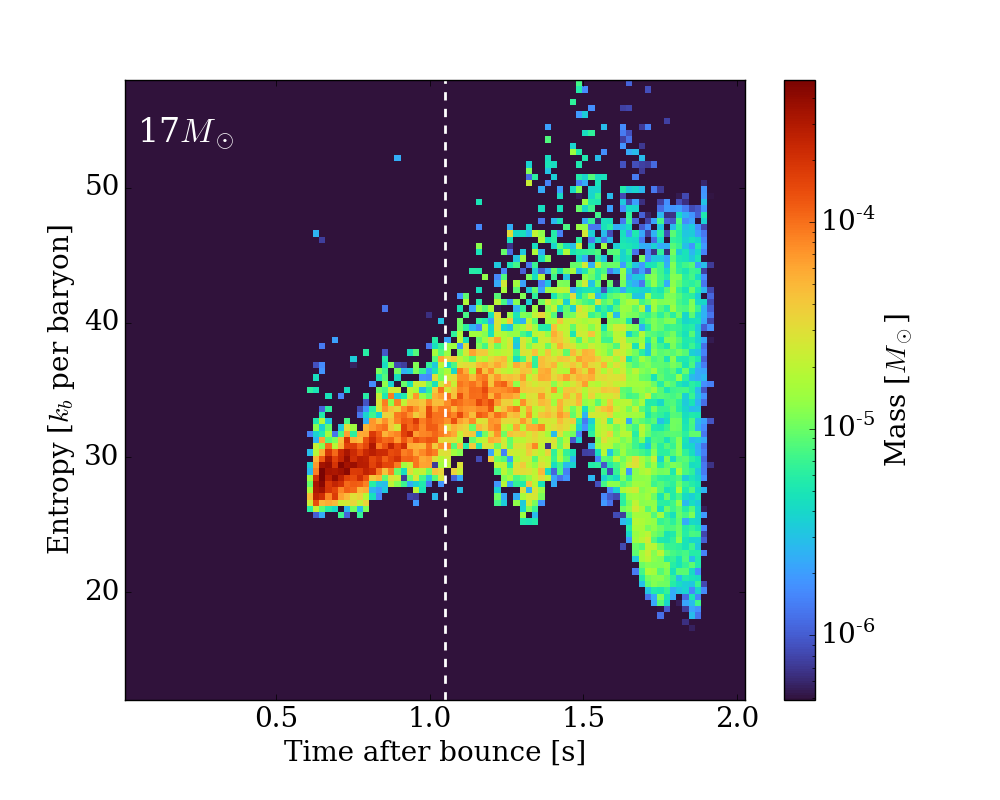}
    \includegraphics[width=0.48\textwidth]{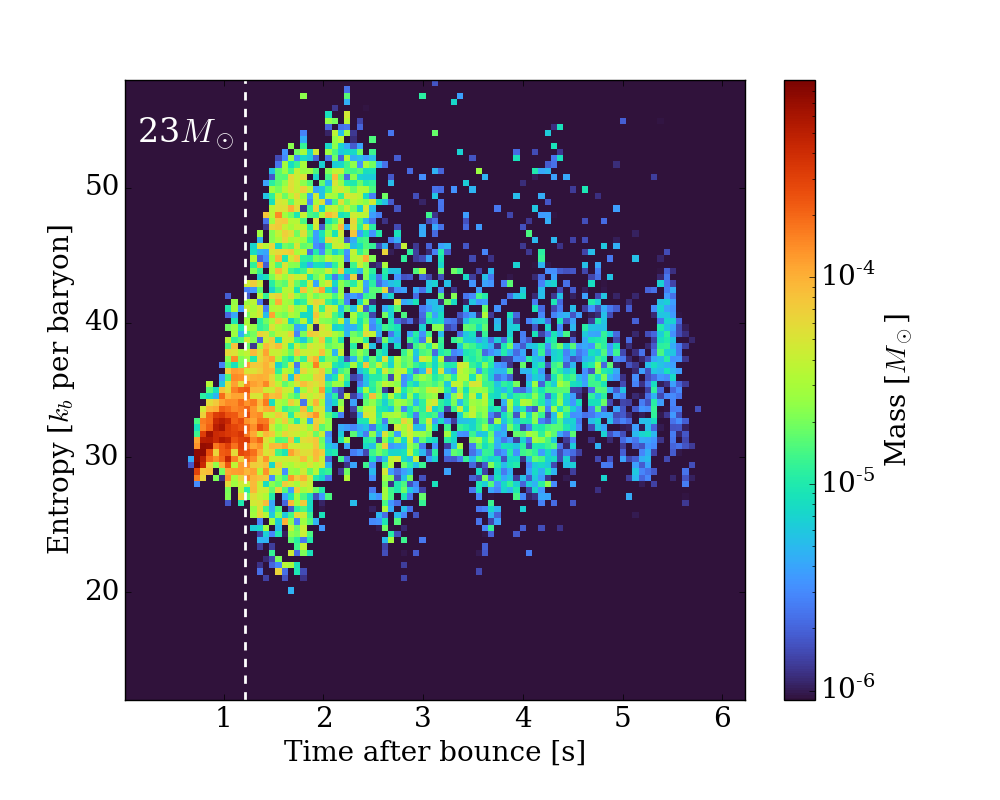}
    \caption{Entropy evolution of the tracer particles during the wind acceleration phase. Color depicts the mass at a given time with that entropy. Both the winds and the early ejecta are included here. The vertical white dashed line indicates the time when the minimum shock radius has reached 3000 km (the fourth wind condition we use in Section \ref{sec:existence}). Lower massive models such as the 9(b), 9.25 and 9.5 $M_\odot$ models show clear increases in the wind entropy. More massive progenitor models do not show such a clear trend. This behavior can be explained by the character of late-time/post-explosion infall. Since these components increase the density in the wind region, the entropy decreases \tianshu{roughly as $S\propto T^3/\rho$}. However, none of our models have wind entropies above 80 $k_b$ per baryon, even if the infall accretion is minimal (e.g., as in the 9(b) model).}
    \label{fig:tracer_t_S}
\end{figure}

Figure \ref{fig:tracer_t_S} depicts the evolution of the entropy of all matter ejected from below 100 km. Matter included here encompasses both the early ejecta and the winds. In the 9(b) $M_\odot$ model, there is a clear transition between two phases of entropy evolution. The first phase in which the entropy grows faster is associated with the early ejecta, while the second phase tracks the predictions of 1D wind solutions (e.g., Figure 14 in \citet{wanajo2023}). Other models also show the two-phase structure, but the entropy evolution in the second phase can be different. The vertical white dashed line indicates the time when the minimum shock radius has reached 3000 km (the fourth wind condition we use in Section \ref{sec:existence}), and matter to the right of this vertical line is identified  with winds. We can see that the time cut we use ensures that the selected tracers represent the wind instead of the early ejecta, but this is a conservative criterion and the early wind phase might be missed in some models.

In Figure \ref{fig:tracer_t_S}, we see that the wind entropy in the less massive models increases with time. This is because accretion in less massive models terminates earlier, resulting in lower densities in the wind region. However, more massive models generally have longer-lasting late-time accretion and in these cases the density in the wind regions don't drop as quickly, leading to more complex behavior in the evolution of the entropy. None of our simulations has a wind entropy above 80 $k_b$ per baryon, which is significantly lower than \tianshu{what} is required for the rapid neutron capture process (r-process).

\begin{figure}
    \centering
    \includegraphics[width=0.48\textwidth]{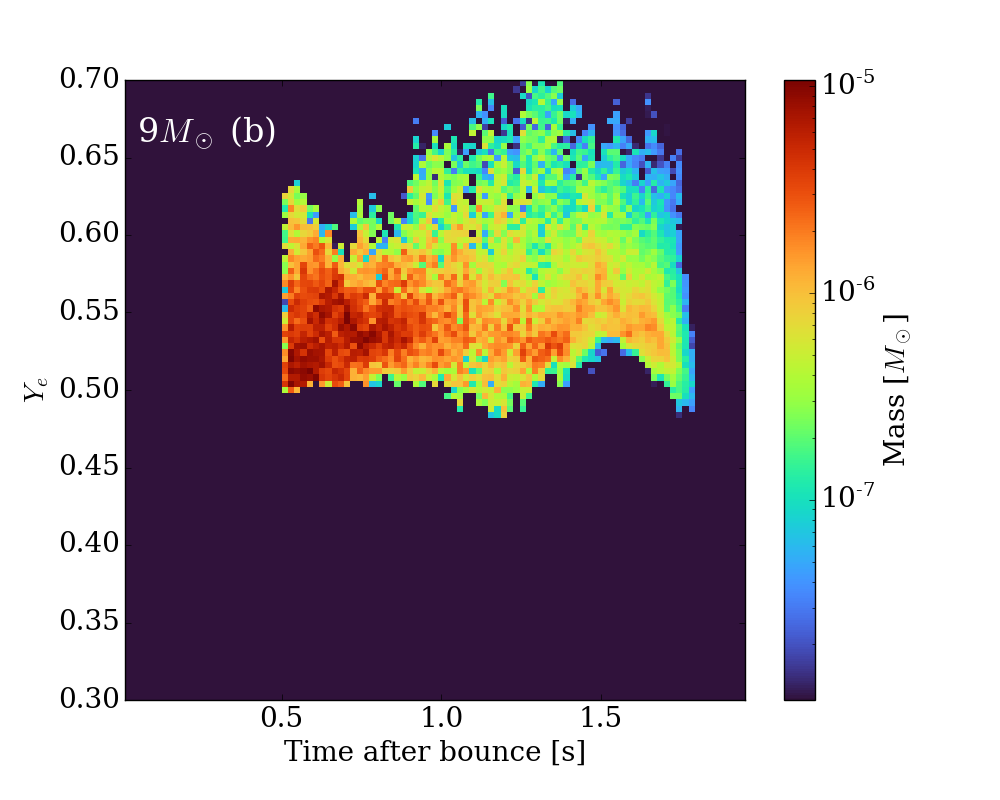}
    \includegraphics[width=0.48\textwidth]{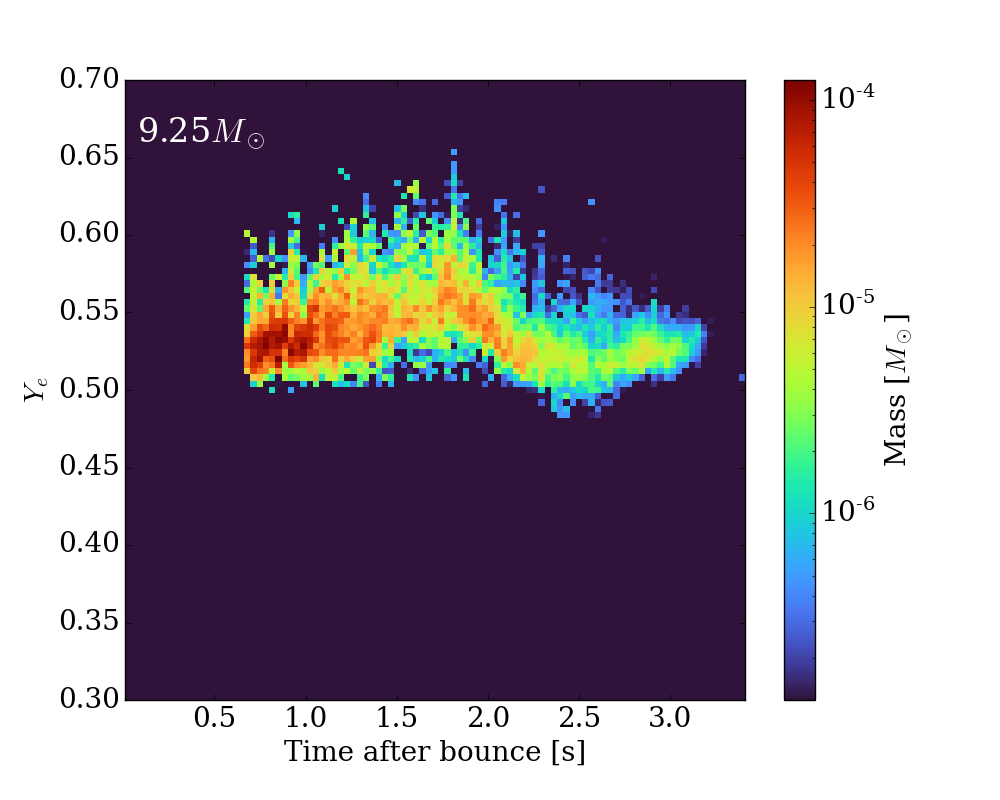}
    \includegraphics[width=0.48\textwidth]{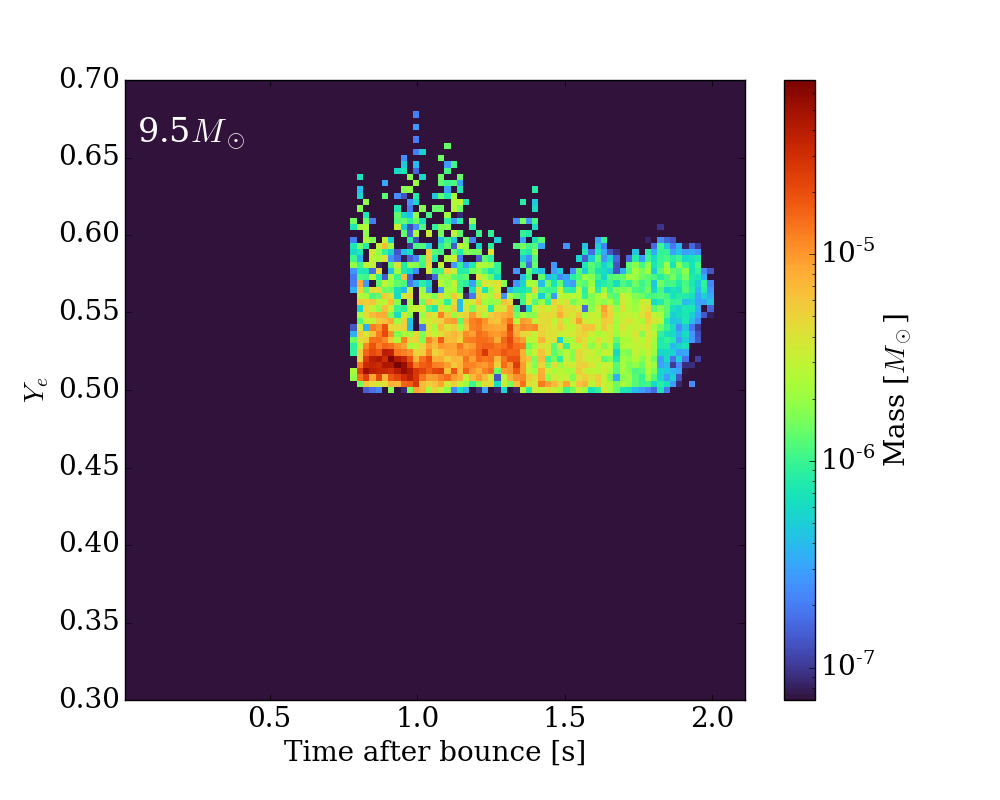}
    \includegraphics[width=0.48\textwidth]{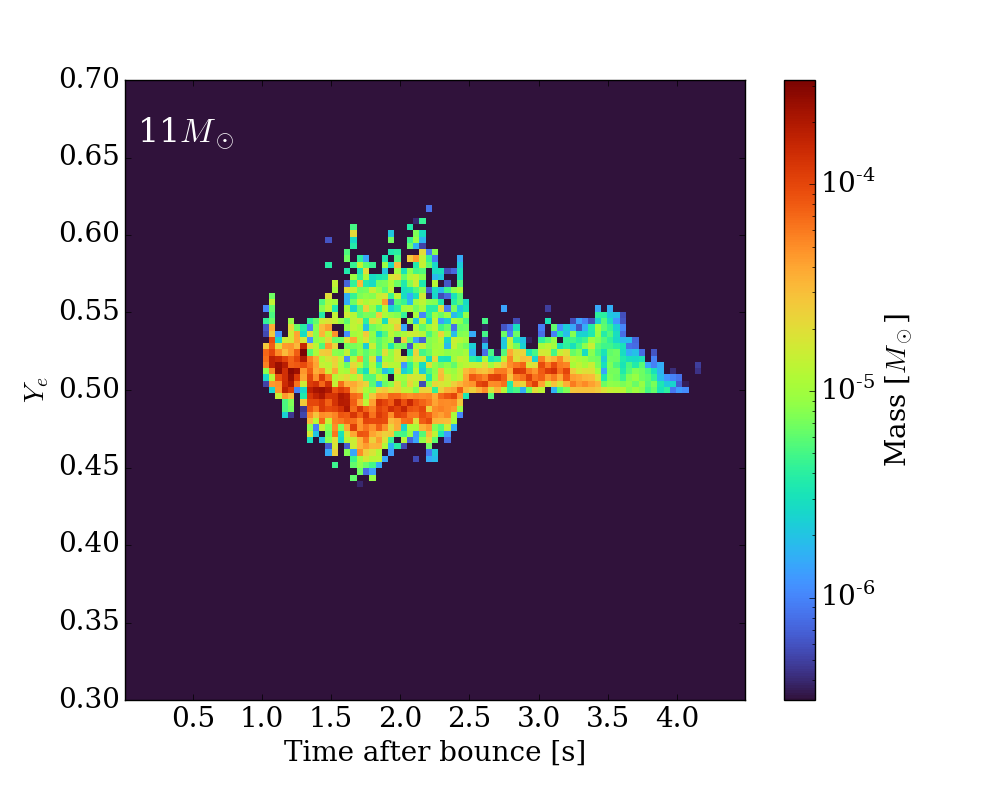}
    \includegraphics[width=0.48\textwidth]{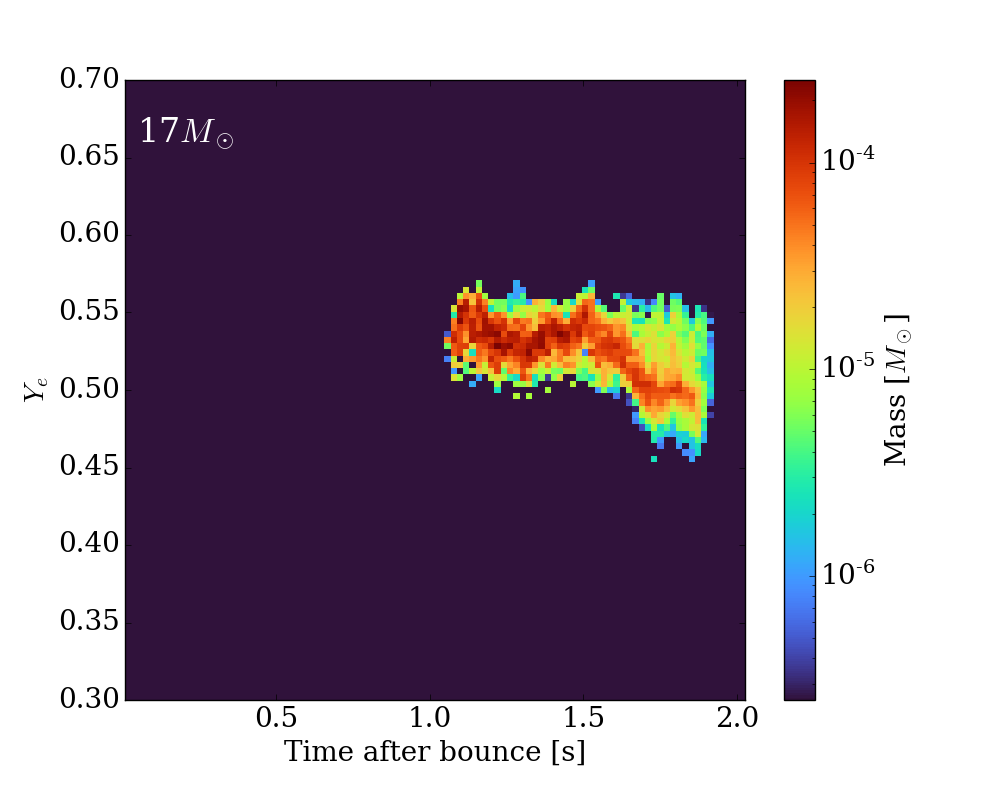}
    \includegraphics[width=0.48\textwidth]{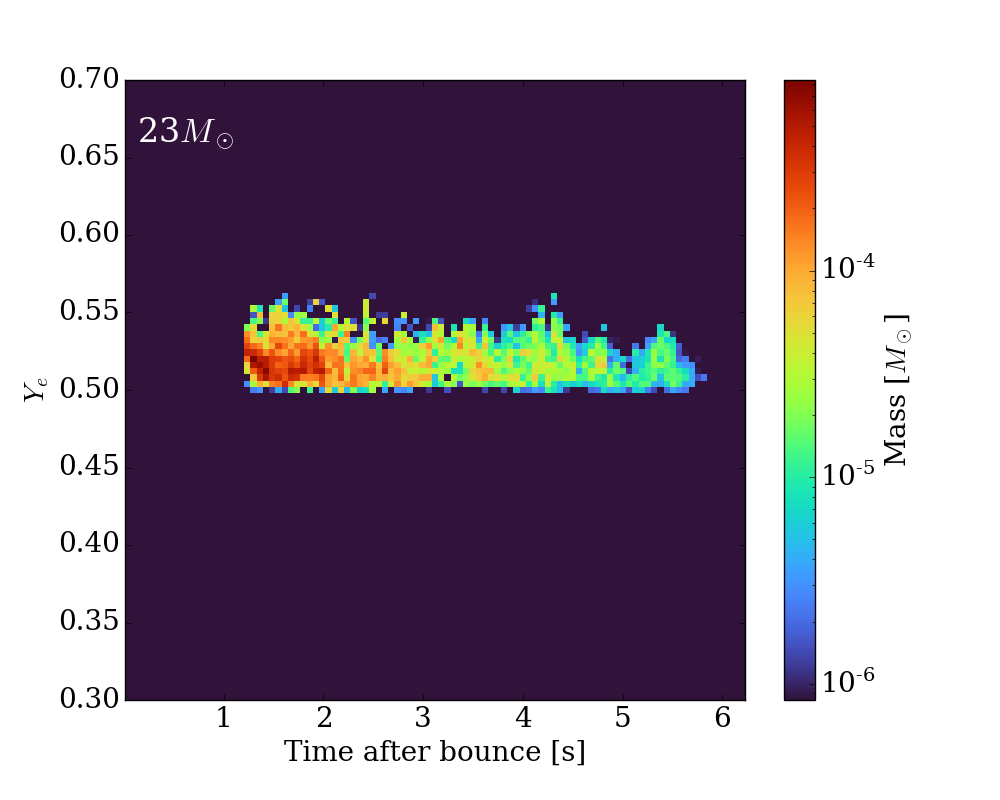}
    \caption{Electron fraction evolution of the tracer particles during the acceleration phase. Color shows the mass at a given time and $Y_e$. Only wind matter is shown here. There is no clear trend in the evolution of $Y_e$. Most wind matter is ejected with $Y_e>0.5$, but there are some neutron-rich ($Y_e<0.5$) phases. This preserves the possibility of weak r-processes. The 9(a) model (not shown here) in particular, featuring imposed initial perturbations and a quick explosion launch, has fractionally more neutron-rich ejecta.  It is interesting that the neutron-rich phases in the 11 and 17 $M_\odot$ models also manifest a fast decrease in entropy. It is possible that strong infall triggers some mixing in the PNS which might inject more neutron-rich material into the wind-forming region.}
    \label{fig:tracer_t_ye}
\end{figure}

The evolution of electron fraction $Y_e$ is shown in Figure \ref{fig:tracer_t_ye}. Similar to Figure \ref{fig:tracer_t_S}, the matter upon which we focus here includes both the early ejecta and the winds, and they are separated by the vertical white dashed line. The $Y_e$ is measured when the material freezes out from nuclear statistical equilibrium (NSE), i.e., when the temperature drops below 0.6 MeV ($\sim$7 GK). Most wind matter is proton-rich, but during some time periods it can be a bit neutron-rich. We don't yet find a clear progenitor-dependent trend in the $Y_e$ temporal evolution. However, it is interesting that the neutron-rich phases in the 11 and 17 $M_\odot$ models also manifest a fast decrease in the entropy. It is possible that strong late-time infall might result in mixing in the outer PNS.  Such mixing may inject neutron-rich matter into the wind-forming region, thereby increasing the density there and resulting in slightly lower $Y_e$s at the wind base. 

\subsection{Nucleosynthesis}\label{sec:nucleosynthesis}
\begin{figure}
    \centering
    \includegraphics[width=0.48\textwidth]{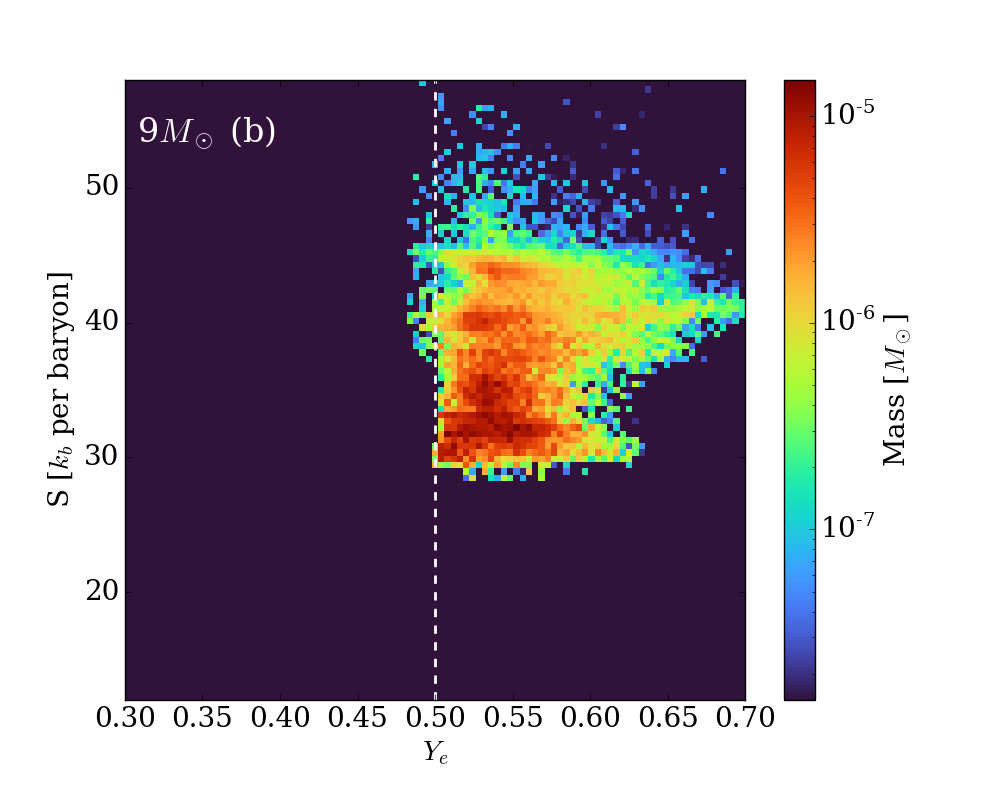}
    \includegraphics[width=0.48\textwidth]{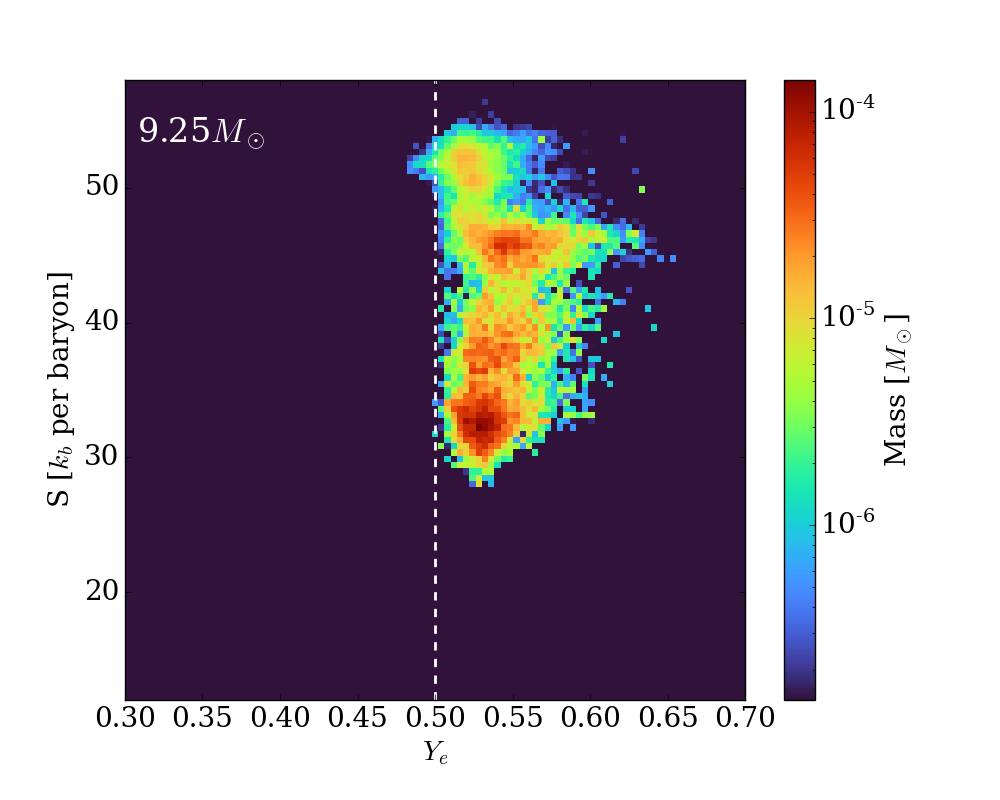}
    \includegraphics[width=0.48\textwidth]{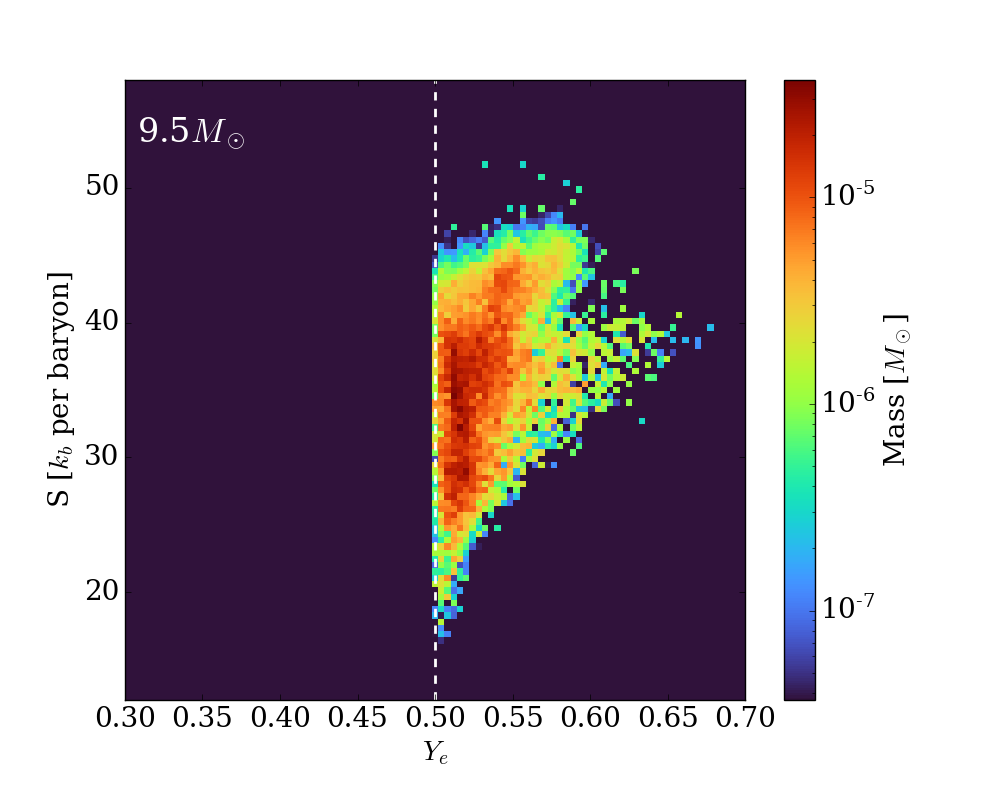}
    \includegraphics[width=0.48\textwidth]{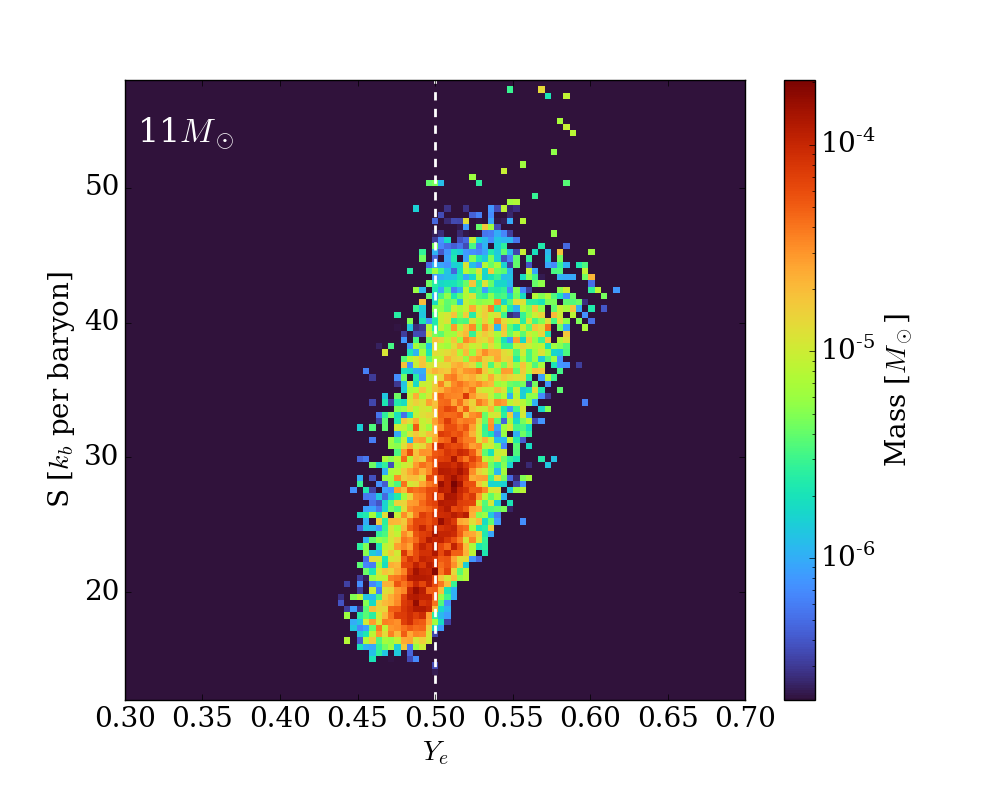}
    \includegraphics[width=0.48\textwidth]{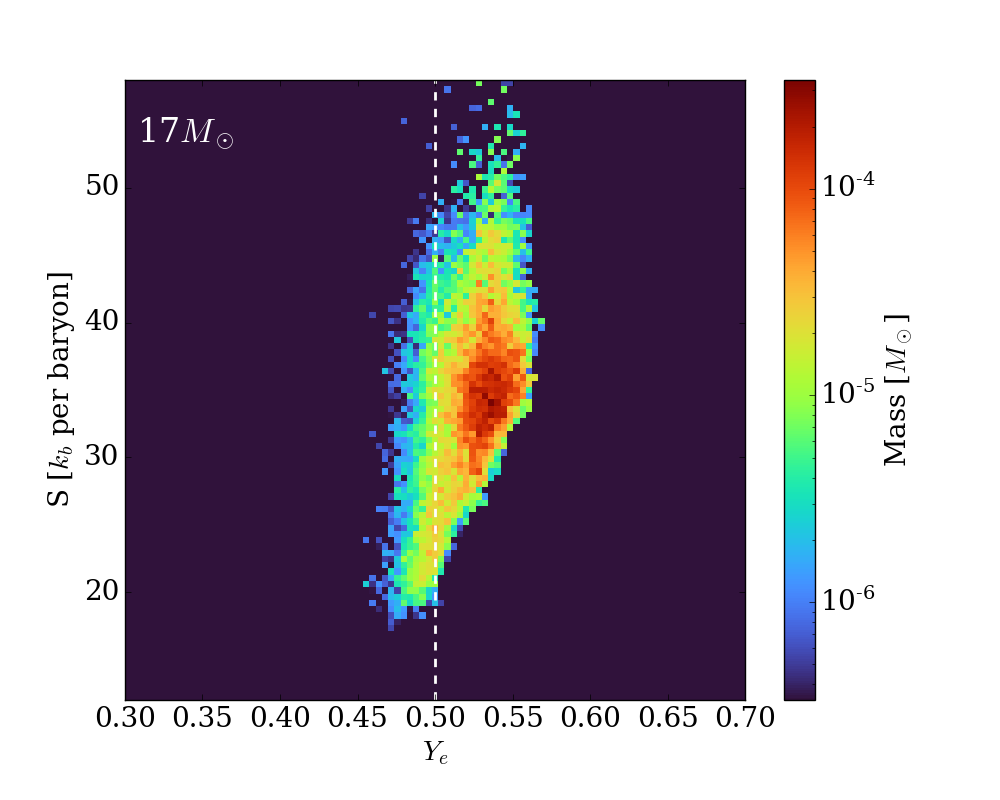}
    \includegraphics[width=0.48\textwidth]{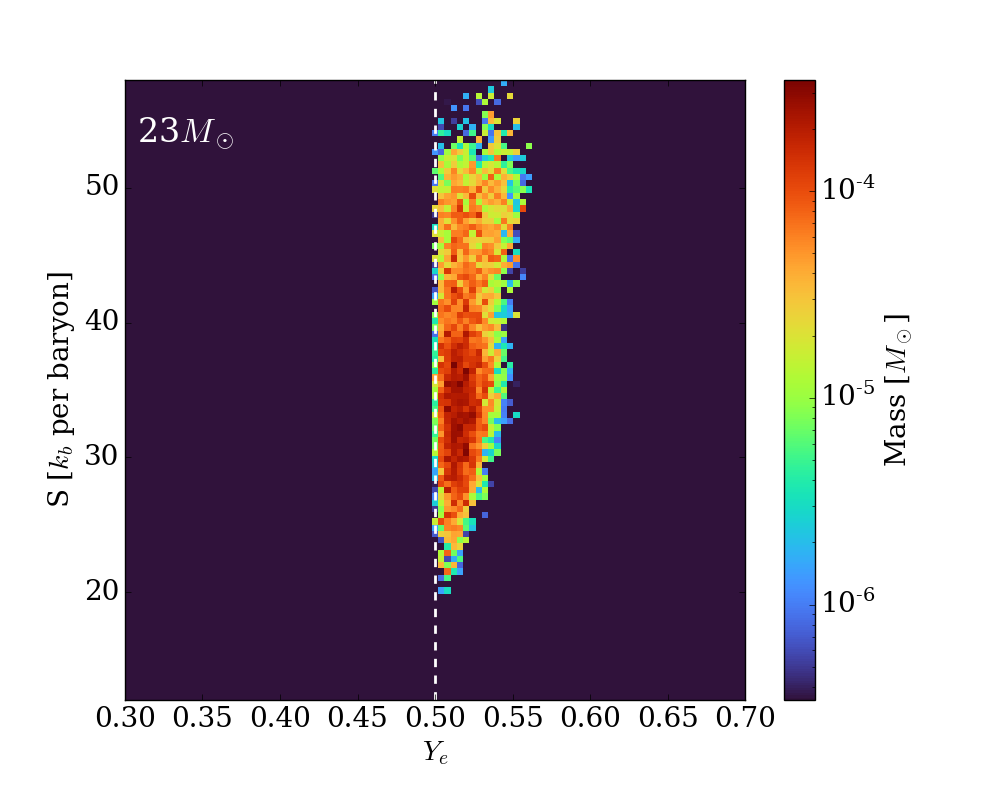}
    \caption{Electron fraction vs. entropy of the tracer particles during the acceleration phase. Color shows the mass at a given $Y_e$ and entropy. $Y_e=0.5$ is marked as a vertical white dashed line. Only wind matter is shown here. The fact that wind matter is at most moderately neutron-rich and does not have high enough entropy eliminates the possibility of a strong r-process. However, a weak r-process is allowed, as expected by \citet{wanajo2013}, \citet{arcones2013}, \citet{wanajo2023}, and seen in this study (see Figure \ref{fig:yield}.)
    }
    \label{fig:tracer_ye_S}
\end{figure}

Figure \ref{fig:tracer_ye_S} shows the entropy and $Y_e$ of the wind. As mentioned in Section \ref{sec:evolution}, the entropy never grows above 80 $k_b$ per baryon and the $Y_e$ can only be slightly below 0.5. This rules out the possibility of strong r-process in the winds. However, it is still possible to produce some of the lightest r-process elements, as discussed in \citet{wanajo2013,arcones2013,wanajo2023}. But the yield of such isotopes can vary a lot due to the large variation in the $Y_e$ evolution of the winds. 

\begin{figure}
    \centering
    \includegraphics[width=0.90\textwidth]{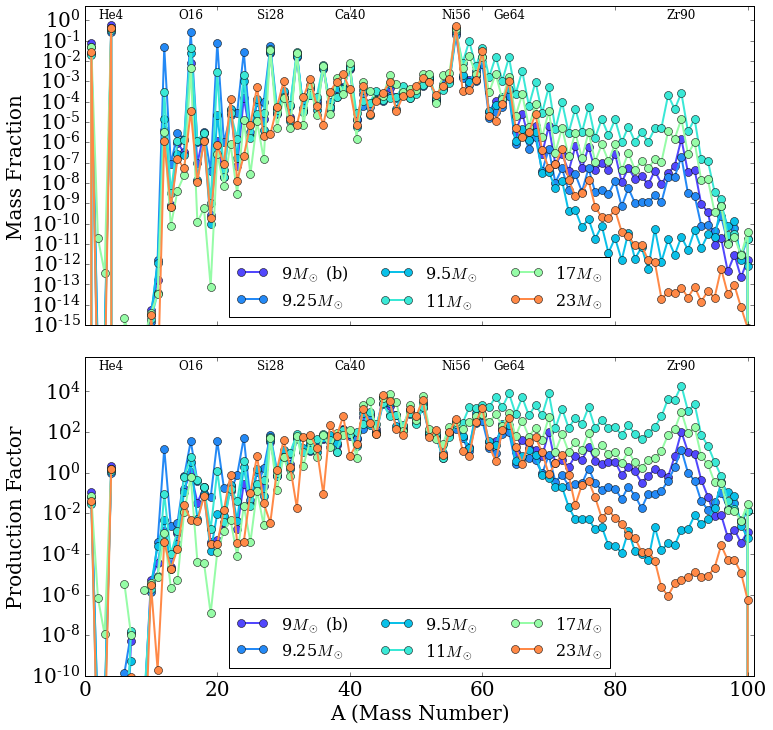}
    \caption{Nucleosynthesis yields and production factors \tianshu{(the ratio of the calculated abundances and solar abundances)} in the winds. Top: Mass fraction at the end of six representative simulations. Bottom: Production factors at the end of those simulations. The production factor is calculated based on the solar system abundances given in \citet{lodders2021}. Isotopes with the same mass number (A) are summed together in both panels. Some interesting isotope positions are indicated at the top of each panel. This figure confirms that the winds are able to generate heavy isotopes up to the $^{90}$Zr peak, thanks to the weak r-process and the $\nu p$-process. However, the 9.5 and 23 $M_\odot$ models don't generate as many heavy elements. One reason is that they have very little neutron-rich ejecta and, therefore, don't experience any significant r-process. Another reason is that the $\nu p$-process (include in this study) can generate heavy elements only if the wind terminates neither too early nor too late \citep{arcones2013}, which is probably not satisfied by these models. This means that the $\nu p$-process will be sensitive to the morphology of the winds and the explosion. We therefore conclude that it is difficult to predict $\nu p$-process yields without doing actual 3D simulations.}
    \label{fig:yield}
\end{figure}

Figure \ref{fig:yield} portrays the bulk nucleosynthesis in the context of our 3D CCSN models as calculated using SkyNet \citep{lippuner2017}. The production factor is calculated based on the solar system abundances in \citet{lodders2021}. In these calculations, we don't distinguish winds that terminate at different radii. As a result, the nucleosynthesis results shown here reflect a range of termination times. The termination time determines how long the material stays above the minimum temperature ($\sim$0.2 MeV) for which most nucleosynthesis occurs. Therefore, winds that terminate earlier create isotopes up to the iron-group via $\alpha$-rich freeze-out and show similar behavior to that seen during classical explosive nucleosynthesis, while the almost freely-expanding matter can have higher neutron-to-seed ratios with which to build elements up to Zr. This component is uniquely associated with winds. In this figure, we see that winds generally have higher helium fractions. For models with more neutron-rich matter (such as the 11 $M_\odot$ model and 9(a)/9(b) models), isotopes up to $^{90}$Zr can be produced via a  weak r-process. The $\nu p$-process (which we include in this study) can also help produce heavy elements, but a strong $\nu p$-process occurs only if the wind terminates neither too early nor too late \citep{arcones2013}, which is a condition probably hard to satisfy in realistic simulations. This means that the $\nu p$-process will be sensitive to the morphology of the winds and the explosion, and that it is hard to predict its yield without doing actual 3D simulations. The 9(a), 9(b), 9.25, 11 and 17 $M_\odot$ models show some production of heavier elements, while the 9.5 and 23 $M_\odot$ models don't. This is in part a direct result of the chaotic $Y_e$ evolution shown in Figure \ref{fig:tracer_t_ye}, and it also indicates that the nucleosynthesis results may depend on the morphology of the winds and the explosion.

Although our simulations automatically include sound waves from the PNS and the heating and momentum flux due to them, we don't see strong r-process predicted in \citet{nevins2023}. There are some possible reasons. First, the entropy in winds is significantly lower than in those 1D simulations. Even if an extra energy source is included, it's unlikely to increase the entropy from below 80 to above a few hundreds of $k_b$ per baryon. Second, the assumed $Y_e=0.48$ in \citet{nevins2023} is rarely achieved by most of our models, except the 11 and 17 $M_\odot$ models. In addition, our simulation resolution may not be high enough to fully capture the non-linear sound-wave damping. 

It is worth noting that the mass of wind matter is much less than the mass of the total ejected matter, even if we don't include the outer envelope swept away by the explosion shock wave. In all our models, the peak mass flow rate in winds is at most a few $10^{-2}$ $M_\odot$s$^{-1}$, and the mass in winds is no more than a few percent of a solar mass (See Table \ref{tab:simulation_summary}). Another important point is that the transition in thermal properties between the early ejecta and the later neutrino-driven wind is smooth, so there is no sudden change in the nucleosynthesis. For the purposes of this paper, we applied a time cut (the fourth condition in Section \ref{sec:existence}) to distinguish the early ejecta from the wind, but they should be considered jointly in a more detailed nucleosynthetic analysis. We leave such a nuanced study to a future paper.

\subsection{Morphology} \label{sec:morphology}
The morphology of the wind regions is determined jointly by the winds, the late-time infall, and the earlier matter blasted outward by the supernova shock. Figure \ref{fig:second-shock} depicts the morphology of high-velocity regions on a 10000-km scale. The larger-scale wind regions are located along directions where the shock radii are larger. Actually, the wind direction coincides with the center of the higher-velocity bubbles (green bubbles in Figure \ref{fig:second-shock}), but not all higher-velocity bubbles have wind buried inside. This is either because some higher-velocity bubbles are not strong enough to clear out the infall and develop a wind, or because the winds inside them have decelerated and merged into the general flow during the expansion.
In addition, there seems to be a correlation between the open angle of the wind region and the explosion energy. Winds in more energetic explosions (like the 17 $M_\odot$ model) tend to have larger opening angles. These two correlations can be explained in a similar way. In directions where there exist relatively higher shock velocities and more energetic ejecta, the winds tend to sweep away the outer matter earlier and more efficiently; as a consequence, the winds develop more easily. 

\begin{figure}
    \centering
    \includegraphics[width=0.48\textwidth]{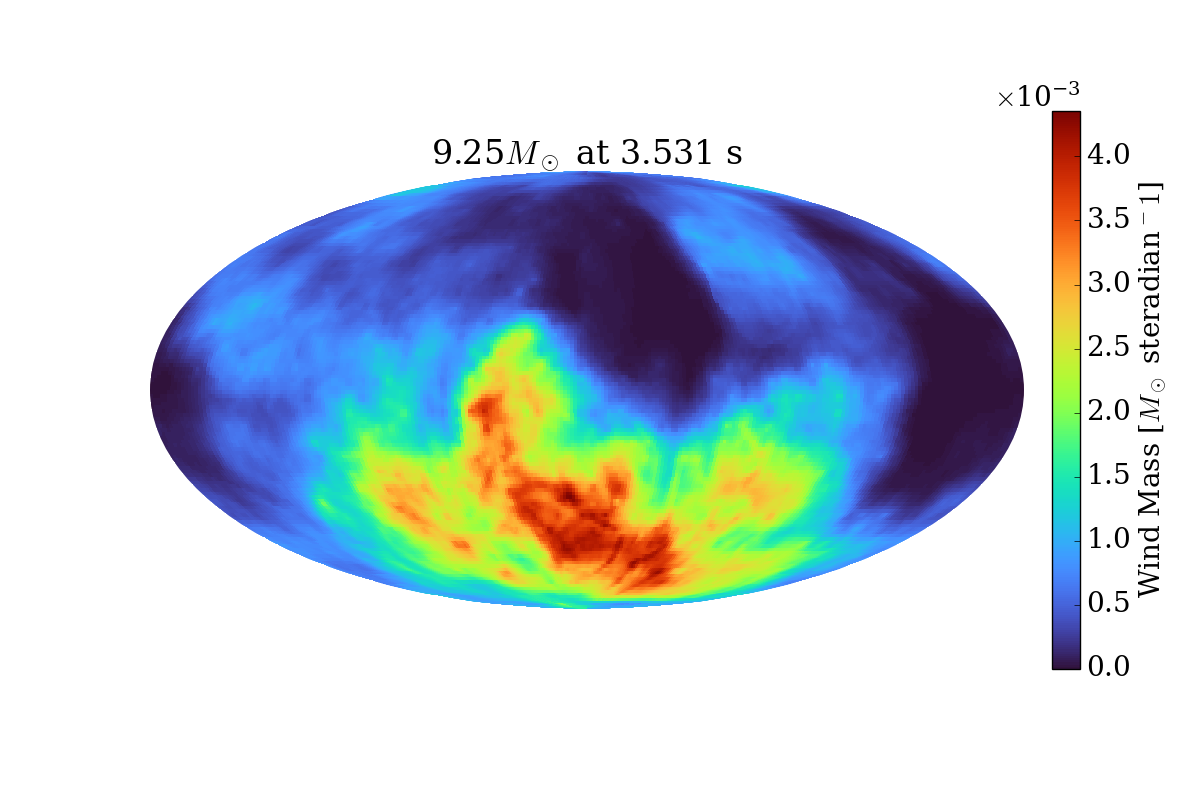}
    \includegraphics[width=0.48\textwidth]{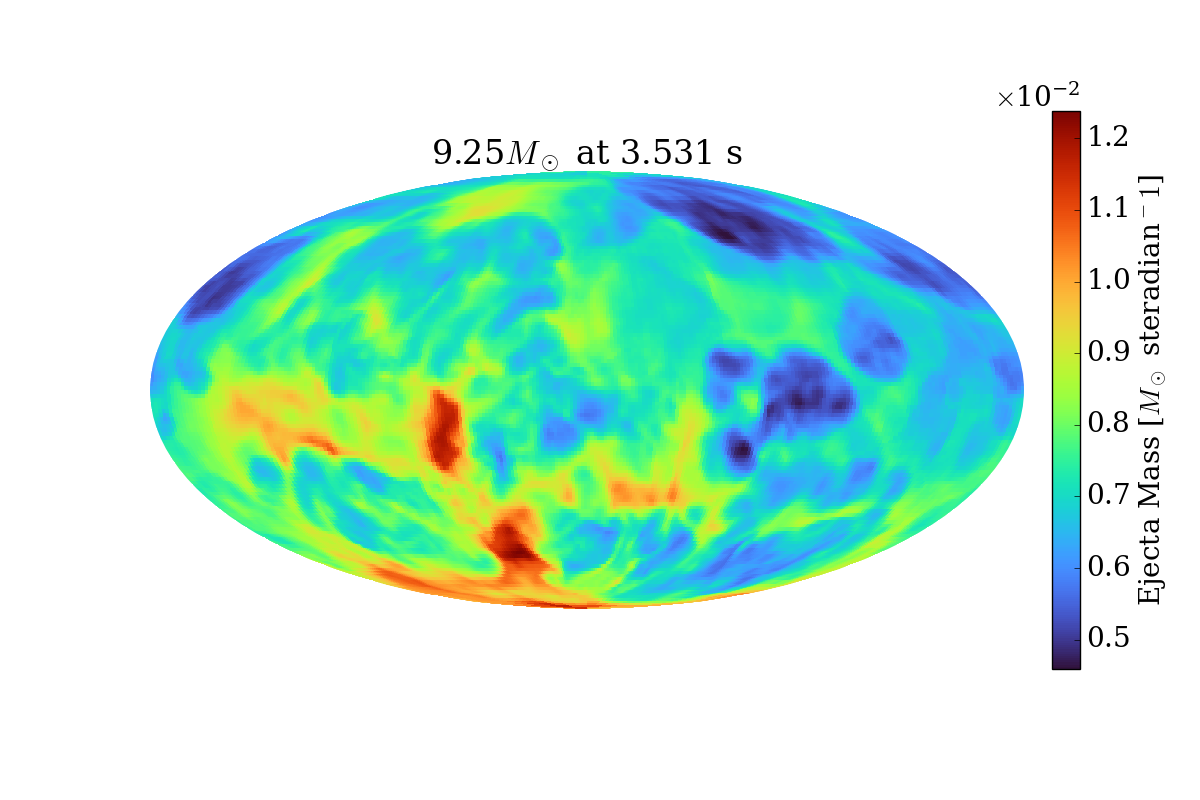}
    \includegraphics[width=0.48\textwidth]{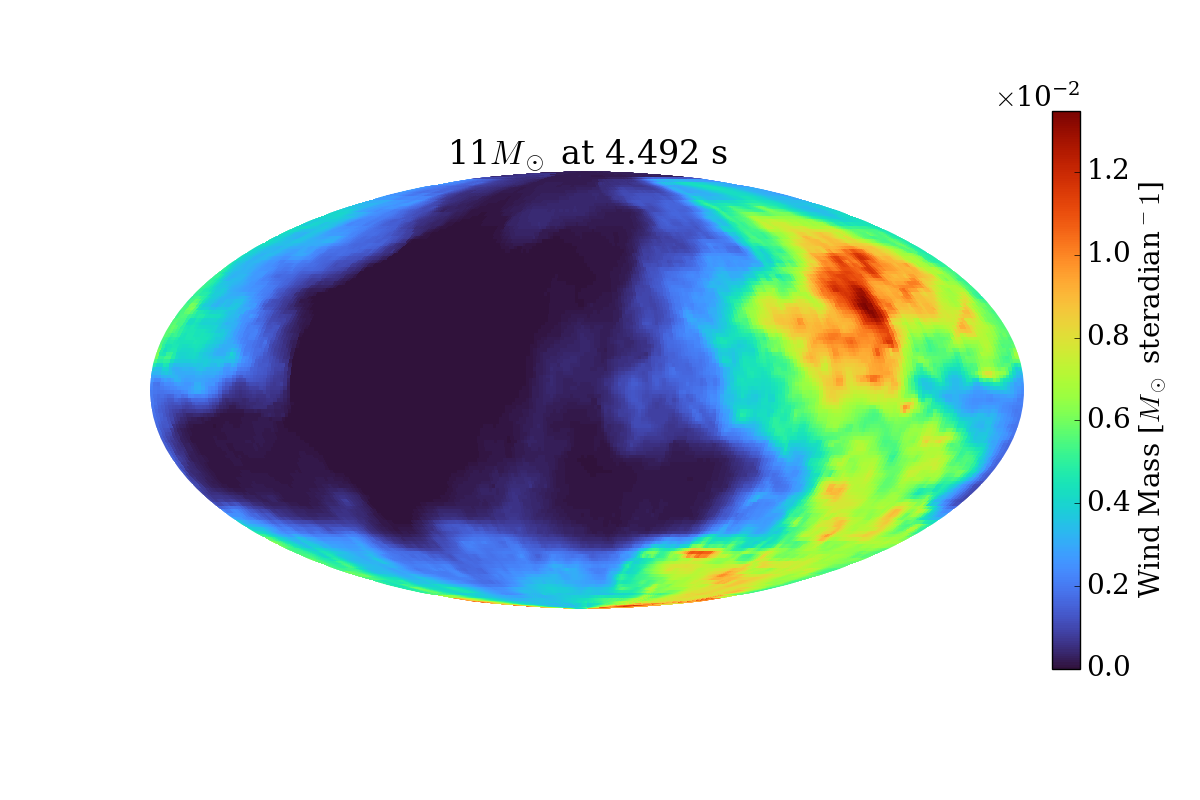}
    \includegraphics[width=0.48\textwidth]{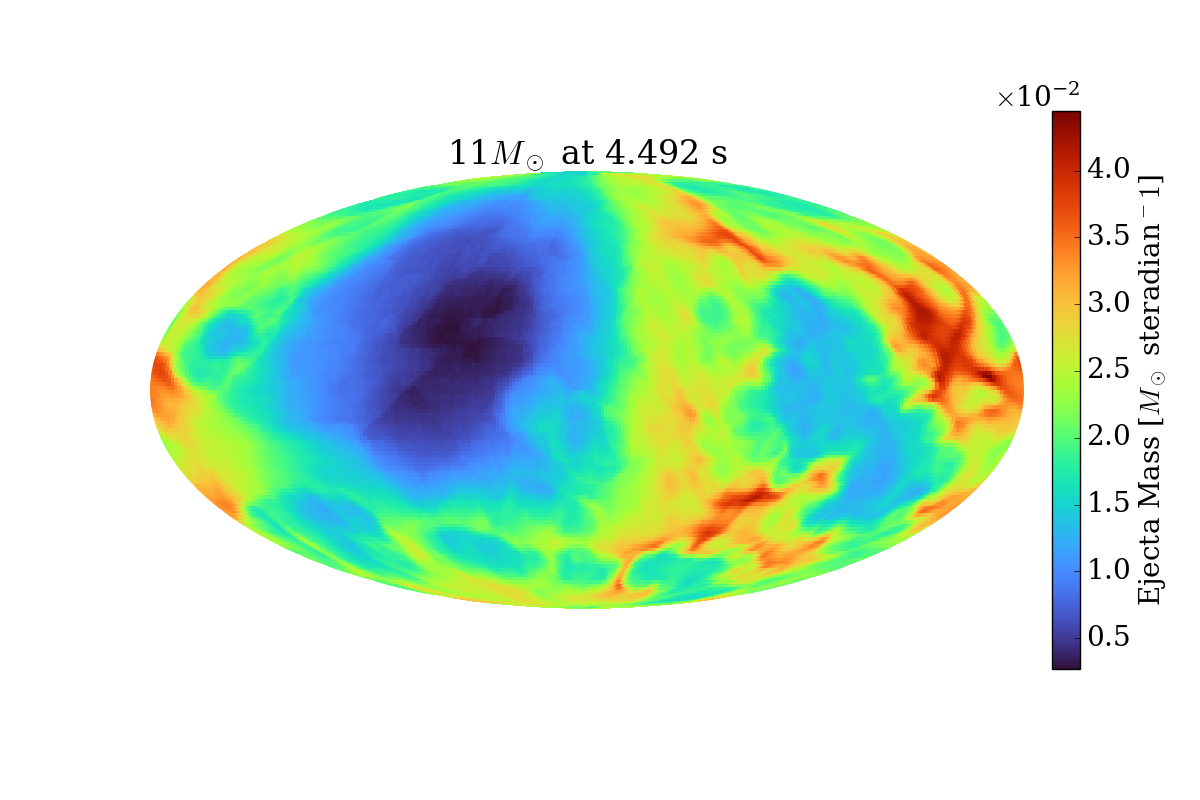}
    \includegraphics[width=0.48\textwidth]{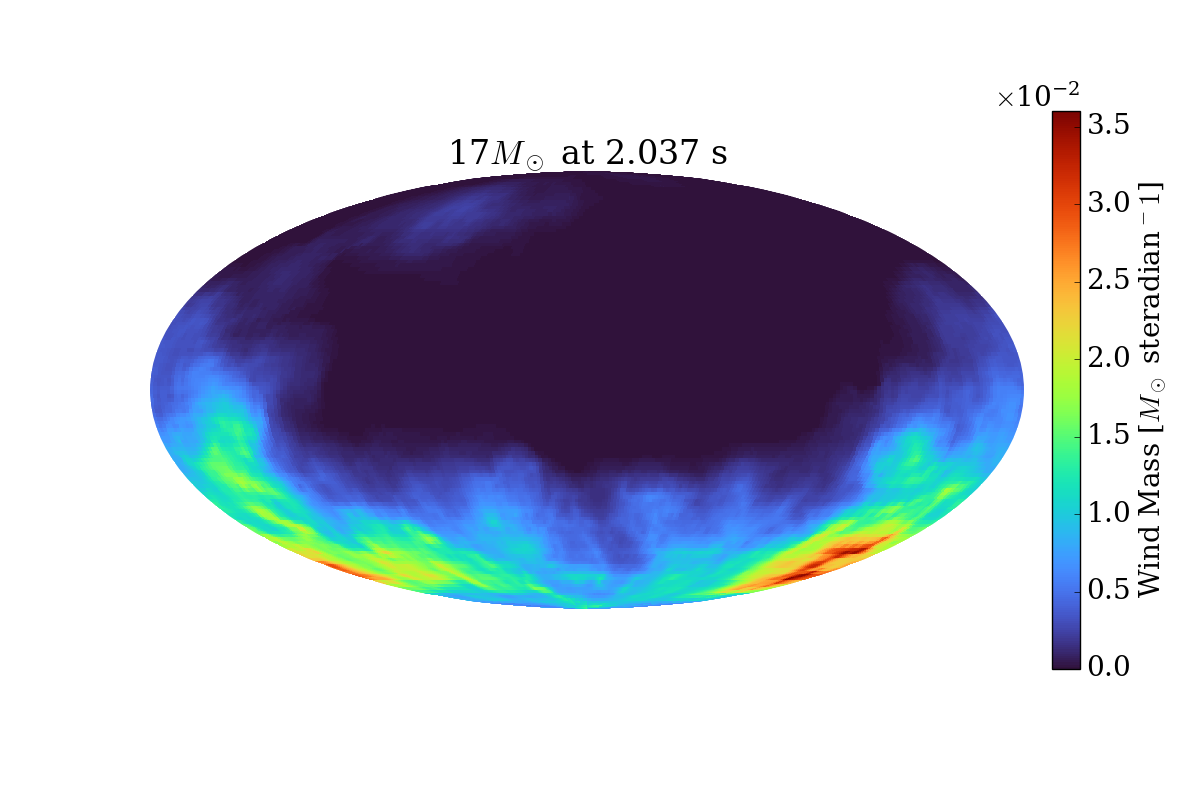}
    \includegraphics[width=0.48\textwidth]{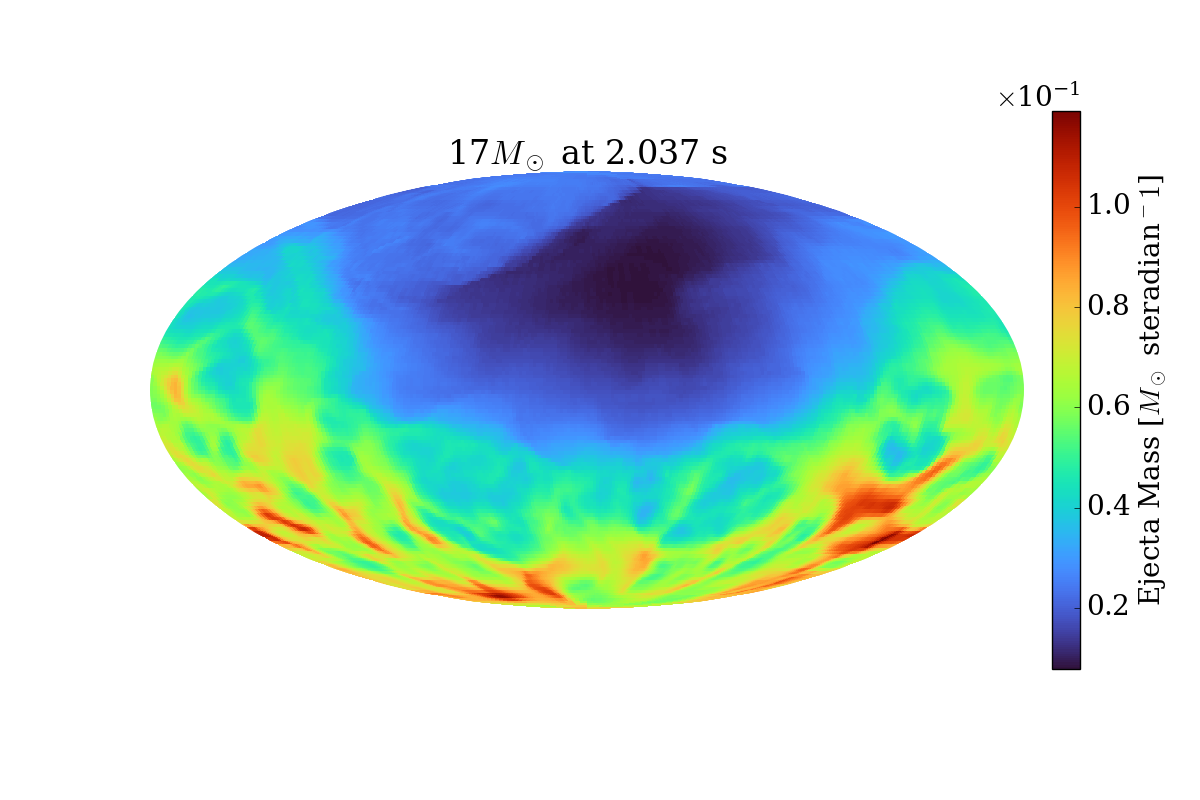}
    \caption{Angular mass distributions of the wind and all ejecta at the end of the simulation for the 9.25, 11, and 17 $M_\odot$ models. Left: wind ejecta. Right: All ejected matter (all matter with a positive binding energy) at the end of the simulations; the outer stellar envelopes not on our grids are not included. It can be seen that on larger angular scales the winds are generally distributed along the same directions as the explosion, while on smaller scales they are much less correlated with the distribution of the total ejecta. This is because the winds are more associated with the high-velocity, high-entropy, low-density bubbles.}
    \label{fig:angular-distribution}
\end{figure}

After hitting the wind-termination shock, the wind matter enters the slowly-moving region and is mixed with other matter. Therefore, the wind region itself doesn't necessarily follow the distribution of wind matter. Figure \ref{fig:angular-distribution} compares the angular mass distribution of the wind matter and all matter with a positive binding energy (all ejecta) at the end of the simulations of the 9.25, 11 and 17 $M_{\odot}$ models. On larger scales (e.g., in the dipolar structures), the wind and ejecta distributions are correlated, since both components emerge more easily along the larger shock radius directions. On smaller scales, the winds and ejecta can be anti-correlated, with winds more distributed in the low-mass directions. This is because the wind matter is found more often in the high-velocity, high-entropy, low-density explosion bubbles.

\section{Conclusions}
\label{sec:conclusion}
In this paper, we have analyzed the properties of early-phase neutrino-driven winds using twelve long-duration 3D state-of-the-art core-collapse simulations covering a large progenitor mass range from 9 to 60 solar masses. This is the first comprehensive paper on winds in the context of sophisticated 3D CCSN simulations. We define the wind to be the transonic outflow that originates below 80 km, and we use a time cut to distinguish the wind and early ejecta launched by the supernova shock wave itself (see Section \ref{sec:definition} and \ref{sec:existence}). This is a technically simple definition which captures most of the central wind features. 

We find that the winds emerge naturally after the successful explosion, and that they universally generate wind-termination shocks (also known as secondary shocks) behind the primary explosion shock wave. The winds are seen in all simulations, indicating that they are a common phenomenon.  However, the winds are generally aspherical, and in more massive (higher compactness) progenitor models they are distorted and channeled by long-lasting infall. 

The velocity profiles of the winds clearly show the transonic feature characteristic of winds. While all models experience similar inaugurating blast phases, winds in more massive models with higher compactness experience larger velocity variations due both to interaction with the primary blast ejecta and to the turbulent velocity field just above the PNS in its atmosphere. We find that there is at most a few percent of a solar mass in the wind component, while the energy carried by the wind to infinity can be as much as 10$-$20\% of the total explosion energy.

Neutrino-driven winds in 3D simulations approximately follow the same $\dot{M}_{wind}\propto L^{2.5}$ relation predicted by the 1D stationary wind solutions. However, the $\dot{M}_{wind}\propto M_{pns}^{-2}$ relation is not seen. Models with higher PNS masses also have stronger wind mass loss rates. This is in part because the higher-PNS mass models have higher neutrino luminosities and longer-lasting infall, which itself brings more matter into the wind-forming region. The entropy evolution of the $9$ $M_\odot$ model follows the 1D predictions very well, while the wind entropy in more massive models increases more slowly and has a greater spread in values. Our models with high PNS masses don't manifest the high entropies often predicted in 1D studies (e.g., \citet{wanajo2023}), and none of our models has an entropy above 80 $k_b$ per baryon at the termination of the simulation. The electron fraction ($Y_e$) evolution in all our models is stochastic, but some wind matter can have $Y_e<0.5$. This allows a weak r-process to occur. In our calculations, the first peak of the r-process up to Zirconium can be synthesized. 

Neutrino-driven winds are more likely to emerge along directions with the highest blast shock velocities, because in those directions the outer matter has been more efficiently cleared away. After hitting the wind-termination shock, wind matter decelerates. Most wind matter resides in the relative high-velocity, high-entropy, and low-density bubbles. Therefore, on larger angular scales the winds are distributed along the same directions as the primary blast ejecta, while on smaller scales the wind matter is not so tightly correlated with those directions, instead concentrating in low-density pockets.

We note that our study, though it encompasses 3D simulations of unprecedented duration after bounce,  is still limited to the first few seconds of the neutrino-driven winds; we have yet to capture the entire PNS wind phase. Moreover, it is technically difficult to distinguish completely the wind from the tail end of the earlier explosion ejecta. We have applied a time cut, but this might eliminate a small fraction of the early phase of the wind. Because the mass flow rate in winds decays quickly with time, missing the early phases may lead to non-negligible differences in the inferred properties of the wind vis \`a vis the summed ejecta. In addition, the tracers we employed in this study were post-processed based on the fluid velocity field saved every millisecond. Although a sub-iteration method was used to increase the temporal resolution, the tracer trajectory may still deviate from the true trajectory in the turbulent region around the PNS. However, this has little influence on the nucleosynthetic yields calculated and how they may be partitioned between the wind and earlier blast components.  This is due in part to the fact that the temperature in the turbulence region is always above our NSE criteria ($\sim$7 GK) and no nucleosynthetic process has then yet started. But since we find that the atmosphere from which the PNS wind emerges is actually turbulent, the simple traditional picture found in the literature of a spherical wind emerging from a quiescent atmosphere is challenged by our new 3D insights into its true character.  

\section*{Acknowledgments}

We thank Matt Coleman and David Vartanyan for their technical help and advice during the conduct of this investigation. We also acknowledge support from the U.~S.\ Department of Energy Office of Science and the Office of Advanced Scientific Computing Research via the Scientific Discovery through Advanced Computing (SciDAC4) program and Grant DE-SC0018297 (subaward 00009650), support from the U.~S.\ National Science Foundation (NSF) under Grants AST-1714267 and PHY-1804048 (the latter via the Max-Planck/Princeton Center (MPPC) for Plasma Physics), and support from NASA under award JWST-GO-01947.011-A.  A generous award of computer time was provided by the INCITE program, using resources of the Argonne Leadership Computing Facility, a DOE Office of Science User Facility supported under Contract DE-AC02-06CH11357. We also acknowledge access to the Frontera cluster (under awards AST20020 and AST21003); this research is part of the Frontera computing project at the Texas Advanced Computing Center \citep{stanzione2020} under NSF award OAC-1818253. In addition, one earlier simulation was performed on Blue Waters under the sustained-petascale computing project, which was supported by the National Science Foundation (awards OCI-0725070 and ACI-1238993) and the state of Illinois. Blue Waters was a joint effort of the University of Illinois at Urbana--Champaign and its National Center for Supercomputing Applications. Finally, the authors acknowledge computational resources provided by the high-performance computer center at Princeton University, which is jointly supported by the Princeton Institute for Computational Science and Engineering (PICSciE) and the Princeton University Office of Information Technology, and our continuing allocation at the National Energy Research Scientific Computing Center (NERSC), which is supported by the Office of Science of the U.~S.\ Department of Energy under contract DE-AC03-76SF00098.

\bibliography{sample631}{}
\bibliographystyle{aasjournal}

\clearpage

\begin{table}
    \centering
    \begin{tabular}{c|cccc}
    ZAMS Mass [$M_\odot$]   &Duration [s] &Average (Max) Shock Velocity$^{\dagger}$ [km s$^{-1}$] &\tianshu{Wind Start Time$^{\dagger\dagger}$ [s]} &Wind Mass [$M_\odot$]\\
    \hline
    9(a)    &1.775        &14000(16000)  &0.462   &0.0029   \\ 
    9(b)    &1.950        &13000(15000)  &0.491   &0.0028   \\ 
    9.25    &3.532        &9000(11000)   &0.668   &0.0129   \\ 
    9.5     &2.375        &8000(11000)   &0.781   &0.0058   \\ 
    11      &4.492        &7000(10000)   &1.019   &0.0311   \\ 
    15.01   &3.137        &6000(9000)    &1.050   &0.0145   \\ 
    17      &2.037        &8000(14000)   &1.051   &0.0289   \\ 
    18      &4.328        &6000(10000)   &1.092   &0.0384   \\ 
    20      &2.579        &9000(12000)   &1.347   &0.0194   \\ 
    23      &6.228        &8000(12000)   &1.218   &0.0471   \\ 
    25      &2.464        &10000(11000)  &1.096   &0.0338   \\ 
    60      &3.300        &8000(11000)   &0.932   &0.0483   \\ 
    \end{tabular}
    \caption{This table summarizes the basic properties of all twelve models.  The difference between the 9(a) and 9(b) models is mostly due to the presence of imposed perturbations in the initial model in model 9(a). This led to a slightly earlier and slightly more vigorous explosion launch. The neutrino-driven wind mass given here is defined using the four conditions described in Section \ref{sec:existence}. 
    \tianshu{$^{\dagger}$: The shock velocities are measured at the end of each simulation.}
    \tianshu{$^{\dagger\dagger}$: The wind start times are based on the wind conditions described in Section \ref{sec:existence}. They are shown as the white vertical lines in Figure \ref{fig:tracer_t_S}.}}
    \label{tab:simulation_summary}
\end{table}

\clearpage

\clearpage

%% For this sample we use BibTeX plus aasjournals.bst to generate the
%% the bibliography. The sample631.bib file was populated from ADS. To
%% get the citations to show in the compiled file do the following:
%%
%% pdflatex sample631.tex
%% bibtext sample631
%% pdflatex sample631.tex
%% pdflatex sample631.tex

%% This command is needed to show the entire author+affiliation list when
%% the collaboration and author truncation commands are used.  It has to
%% go at the end of the manuscript.
%\allauthors

%% Include this line if you are using the \added, \replaced, \deleted
%% commands to see a summary list of all changes at the end of the article.
%\listofchanges

\end{document}